\documentclass{emulateapj}
\newcommand{\EMint}{\ensuremath{\int \Ne^2 \dl}}

\newcommand{\Halpha}{H$\alpha$}

\newcommand{\Kalpha}{K$\alpha$}

\newcommand{\Lyalpha}{Ly$\alpha$}

\newcommand{\Ne}{\ensuremath{n_{\mathrm{e}}}}

\newcommand{\NH}{\ensuremath{N_{\mathrm{H}}}}

\newcommand{\angstrom}{\ensuremath{\mbox{\AA}}}
\newcommand{\nm}{\ensuremath{\mbox{\nm}}}
\newcommand{\cm}{\ensuremath{\mbox{cm}}}

\newcommand{\pc}{\ensuremath{\mbox{pc}}}
\newcommand{\kpc}{\ensuremath{\mbox{kpc}}}

\newcommand{\s}{\ensuremath{\mbox{s}}}

\newcommand{\yr}{\ensuremath{\mbox{yr}}}
\newcommand{\Myr}{\ensuremath{\mbox{Myr}}}

\newcommand{\kev}{\ensuremath{\mbox{keV}}}

\newcommand{\erg}{\ensuremath{\mbox{erg}}}

\newcommand{\sr}{\ensuremath{\mbox{sr}}}

\newcommand{\microgauss}{\ensuremath{\mu\mbox{G}}}

\newcommand{\MJy}{\ensuremath{\mbox{MJy}}}

\newcommand{\K}{\ensuremath{\mbox{K}}}

\newcommand{\ph}{\ensuremath{\mbox{photons}}}
\newcommand{\counts}{\ensuremath{\mbox{counts}}}

\newcommand{\parcminsq}{\ensuremath{\mbox{arcmin}^{-2}}}
\newcommand{\pcc}{\ensuremath{\cm^{-3}}}
\newcommand{\pcmsq}{\ensuremath{\cm^{-2}}}

\newcommand{\ps}{\ensuremath{\s^{-1}}}
\newcommand{\psr}{\ensuremath{\sr^{-1}}}

\newcommand{\emismeas}{\ensuremath{\cm^{-6}}\ \pc}

\newcommand{\lineunit}{\ph\ \pcmsq\ \ps\ \psr}

\newcommand{\LU}{\ensuremath{\mbox{L.U.}}}

\newcommand{\pownorm}{\ph\ \pcmsq\ \ps\ \psr\ \ensuremath{\kev^{-1}}}

\newcommand{\presalt}{\pcc\ \K}
\newcommand{\rassrate}{\counts\ \ps\ \parcminsq}

\newcommand{\HI}{H~\textsc{i}}

\newcommand{\NV}{N~\textsc{v}}

\newcommand{\OVI}{O~\textsc{vi}}
\newcommand{\OVII}{O~\textsc{vii}}
\newcommand{\OVIII}{O~\textsc{viii}}

\newcommand{\ace}{\textit{ACE}}

\newcommand{\chandra}{\textit{Chandra}}

\newcommand{\fuse}{\textit{FUSE}}

\newcommand{\iras}{\textit{IRAS}}

\newcommand{\rosat}{\textit{ROSAT}}

\newcommand{\suzaku}{\textit{Suzaku}}

\newcommand{\wind}{\textit{Wind}}
\newcommand{\xmm}{\textit{XMM-Newton}}

\newcommand{\citepossessive}[1]{\citeauthor{#1}'s \citeyearpar{#1}}

\newcommand{\dl}{\ensuremath{\mathrm{d}l}}

\newcommand{\chisq}{\ensuremath{\chi^2}}

\newcommand{\raymondsmith}{\citeauthor{raymond77} (\citeyear{raymond77} and updates)}

\newcommand{\Beff}{\ensuremath{B_\mathrm{eff}}}

\shorttitle{EXTRAPLANAR SUPERNOVA REMNANT}
\shortauthors{HENLEY AND SHELTON}

\begin{document}

\title{A Possible Supernova Remnant high above the Galactic Disk}
\author{David B. Henley and Robin L. Shelton}
\affil{Department of Physics and Astronomy, University of Georgia, Athens, GA 30602}
\email{dbh@physast.uga.edu}

\begin{abstract}
We present the analysis of three \suzaku\ observations of a bright arc in the \rosat\ All-Sky Survey
1/4~\kev\ maps at $l \approx 247\degr$, $b \approx -64\degr$. In particular, we have tested the
hypothesis that the arc is the edge of a bubble blown by an extraplanar supernova. One pointing
direction is near the brightest part of the arc, one is toward the interior of the hypothesized
bubble, and one is toward the bubble exterior.
We fit spectral models generated from 1-D hydrodynamical simulations of extraplanar supernova
remnants (SNRs) to the spectra. The spectra and the size of the arc ($\mathrm{radius} \approx
5\degr$) are reasonably well explained by a model in which the arc is the bright edge of a
$\sim$100,000-yr old SNR located $\sim$1--2~\kpc\ above the disk.  The agreement between the model and the
observations can be improved if the metallicity of the X-ray--emitting gas is $\sim$1/3 solar, which
is plausible, as the dust which sequesters some metals is unlikely to have been destroyed in
the lifetime of the SNR. The width of the arc is larger than that predicted by our SNR model;
this discrepancy is also seen with the Vela SNR, and may be due to the 1-D nature of our
simulations.
If the arc is indeed the edge of an extraplanar SNR, this work supports the idea that extraplanar supernovae
contribute to the heating of the $\sim$million-degree gas in the halo.
\end{abstract}

\keywords{Galaxy: halo --- ISM: bubbles --- ISM: supernova remnants --- X-rays: diffuse background --- X-rays: ISM}


\section{INTRODUCTION}
\label{sec:Introduction}

The discovery of shadows in the 1/4~\kev\ soft X-ray background (SXRB) with \rosat\ demonstrated that there is
$\sim$million-degree gas beyond the Galactic disk, in the Galactic halo \citep{burrows91,snowden91}.  Subsequent
analysis of data from the \rosat All-Sky Survey (RASS; \citealp{snowden97}), and of spectra from pointed observations
with \xmm\ and \suzaku, has confirmed the existence of this hot halo gas. Assuming that the gas is in collisional
ionization equilibrium (CIE), its temperature is $\sim$1--$3 \times 10^6$~K
\citep{snowden98,kuntz00,smith07a,galeazzi07,henley08a,lei09}. High-resolution X-ray absorption line spectroscopy with
the \chandra\ gratings also detected hot halo gas \citep[e.g.][]{yao05}. However, despite nearly 20 years of study,
a fundamental question about the halo remains: how did the hot gas get there?

Various mechanisms may be contributing to the hot halo gas. One possibility is that the hot gas
originated in the disk, heated by stellar winds and supernovae (SNe), and was transferred to the
halo via fountains or chimneys \citep[e.g.,][]{shapiro76,norman89}. Another possible source of hot
halo gas is gravitational heating of infalling intergalactic material, predicted by simulations of
disk galaxy formation \citep[e.g.][]{toft02,rasmussen09}.  A third possibility, which is the subject
of this study, is that the gas is heated \textit{in situ} by SNe above the Galactic disk
\citep{shelton06}. X-ray spectroscopy is the key to distinguishing between these scenarios. For
example, gas that has recently been heated by SNe will be underionized and gas that was heated by
SNe in the distant past will be overionized \citep[e.g.,][]{shelton99}, while gas that has rapidly
expanded out of the disk into the halo will be drastically overionized and recombining
\citep{breitschwerdt94}. Also, gas of extragalactic origin may have a different abundance pattern
from gas of Galactic origin.

\citet{shelton06} considered the soft X-ray emission from an ensemble of isolated supernova remnants 
(SNRs) of different ages and at different heights above the Galactic plane, taking into account the
variation of the SN rate and ambient density as a function of height. She compared this
emission to the 1/4~\kev\ (R12) halo emission determined from the RASS. \citet{snowden98}
decomposed the observed 1/4~\kev\ emission into a foreground component due to the Local Bubble (LB,
a cavity in the interstellar medium of radius $\sim$100~\pc\ in which the Sun resides, thought to be
filled with $\sim$$10^6~\K$ gas), and a distant component, assumed to originate beyond the majority of the
Galaxy's \HI. This distant component is a combination of halo emission and the extragalactic
background. Because the northern Galactic hemisphere contains such anomalous features as the North
Polar Spur, and because at low latitudes one cannot clearly see the halo, \citet{shelton06}
concentrated on high latitudes in the southern Galactic hemisphere. The average de-absorbed count-rate of the
distant component for $b < -65\degr$ is $\sim$$800 \times 10^{-6}$~R12 \rassrate. Subtracting 400~R12 \rassrate\ for
the extragalactic background
\citep[e.g.][]{snowden98} leaves $\sim$$400 \times 10^{-6}$~R12 \rassrate\ for the halo. \citet{shelton06} found that
up to $\sim$80\%\ of this 1/4~\kev\ flux could be explained by a population of extraplanar SNRs. In
this scenario, the vast majority of the flux comes from old SNRs, covering $\sim$30--90\%\ of the
high-Galactic-latitude sky (the fact that they do not cover the whole sky explains the mottled
appearance of the 1/4~\kev\ halo emission in the maps of \citealp{snowden98}). However, the individual old SNRs
are rather dim, and so would be difficult to identify.

\begin{figure*}
\plotone{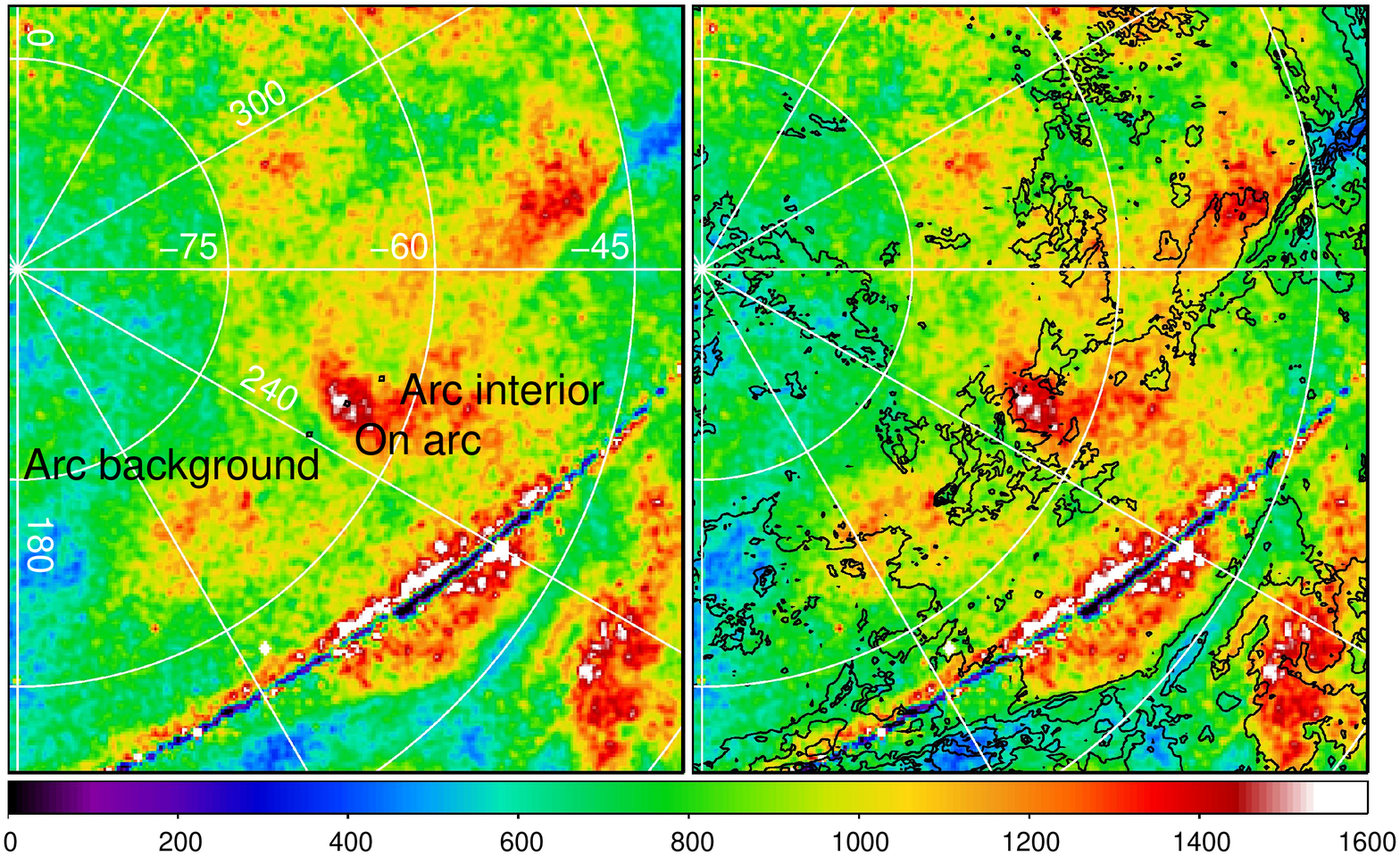}
\caption{\rosat\ All-Sky Survey 1/4~\kev\ maps of the southern Galactic hemisphere \citep{snowden97},
showing the arc that is the subject of this paper. Both panels show the same RASS data, which
have been smoothed with a Gaussian whose standard deviation is 2 times the pixel size. The units on
the color bar are $10^{-6}~\rassrate$. The coordinate grid shows Galactic coordinates.
The small squares in the left panel show the \suzaku\ XIS field of view ($17.8\arcmin \times 17.8\arcmin$)
for our three pointings (labeled ``Arc interior'', ``On arc'',
and ``Arc background''). The contours in the right panel indicate the DIRBE-corrected \iras\ 100-\micron\
intensity \citep{schlegel98}. The contours are at 1, 2, 3, 4, and 5~\MJy~\psr.
\label{fig:ArcImage}}
\end{figure*}

\citet{shelton06} also calculated that $\sim$1\%\ of the high-latitude sky would be
covered by young, bright remnants, which should be easier to identify. (Note that in this context,
``young'' means remnants that are still in the adiabatic phase -- their ages could be up to
$\sim$$10^5$~yr.) This fraction implies that $\sim$1 young, bright extraplanar remnant is expected
per Galactic hemisphere above $|b| \sim 50\degr$. SNRs have been found several hundred pc above the
plane at low latitudes (e.g., SN~1006 at $z \sim 550~\pc$; \citealt{winkler03}), and
\citet{shelton07} and \citet{lei09} have suggested that the X-ray and UV emission from an X-ray--bright
region behind a nearby shadowing filament at $l \approx 279\degr$, $b \approx -47\degr$ is consistent with
the emission from a young remnant. However, to date, no isolated high-latitude extraplanar remnants
have been confirmed. Here we examine another promising candidate for a young
extraplanar remnant, suggested by \citet{shelton06}: a bright arc in the RASS 1/4~\kev\ maps
\citep{snowden97} at $l \approx 247\degr$, $b \approx -64\degr$. This arc is the brightest arc-like
feature in the southern Galactic hemisphere below $\sim$$-40\degr$, and is shown in Figure~\ref{fig:ArcImage}.
The shape of the arc is not due to absorption by intervening material. The right panel of
Figure~\ref{fig:ArcImage} shows the 1/4~\kev\ data overlaid with contours showing the
DIRBE-corrected \iras\ 100-\micron\ intensity \citep{schlegel98}, which traces cool, absorbing
material. One can see that there is increased 100-\micron\ emission (and hence increased absorption
of the background X-rays) to the lower left of the arc. However, the absorbing material does not
completely follow the edge of the arc, implying that the arc's shape is not due to absorption.

Although the RASS maps clearly show the presence of the arc, \rosat's low spectral resolution
($E / \Delta E \sim 1$--3; \citealt{snowden97}) makes detailed studies difficult. Therefore, we have
used the X-ray Imaging Spectrometer (XIS; $E / \Delta E \sim 20$ at $E \sim 1~\kev$;
\citealp{koyama07}) onboard \suzaku\ \citep{mitsuda07} to obtain higher-resolution spectra of the
arc and its surroundings, which we used to test \citepossessive{shelton06} suggestion that the arc
is the edge of a young extraplanar SNR. Our three observing directions are shown in
the left panel of Figure~\ref{fig:ArcImage}.  One observation direction is near the brightest part
of the arc (``On arc''), and one is toward the interior of the hypothesized SNR (``Arc
interior''). The third observing direction (``Arc background'') is off the arc, outside the
hypothesized SNR, and is intended to measure the ambient SXRB in the vicinity of the arc. Note that
the XIS field of view ($17.8\arcmin \times 17.8\arcmin$) is much smaller than the radius of the arc
($\sim$5\degr), so we cannot obtain data from the entire arc region.

The remainder of this paper is organized as follows. In \S\ref{sec:Observations} we describe the
\suzaku\ data reduction and spectral extraction. Our first goal is to establish if there are
differences in the halo component for our three observation directions. We investigate this question
using CIE models, as described in \S\ref{sec:CIEModels}. We find that the halo is brighter and
slightly cooler in the on-arc direction, compared with the other two directions. This CIE analysis
provides a benchmark against which to compare our subsequent analysis, using spectral models based
on \citepossessive{shelton06} hydrodynamical simulations of extraplanar SNRs. This subsequent
analysis, described in \S\ref{sec:SNRModels}, directly addresses the question of whether or not the
arc is the edge of a extraplanar remnant. In addition, if the arc is the edge of a remnant, our
analysis places constraints on the ambient density (corresponding to the height above the disk) and
the age of the remnant. We discuss our results in \S\ref{sec:Discussion}. In particular, we assess
the SNR hypothesis on the basis of the spectrum, brightness, and gross morphology of the arc
(\S\ref{subsec:IsTheArcAnSNR}).  We finish with a summary in \S\ref{sec:Summary}.  Throughout we
quote $1\sigma$ errors.


\section{\textit{SUZAKU} DATA REDUCTION}
\label{sec:Observations}

\subsection{Initial Data Processing and Cleaning}

Table~\ref{tab:Observations} shows the details of our \suzaku\ arc observations.  Our data were
initially processed at NASA Goddard Space Flight Center (GSFC) using version 2 processing,
specifically version 2.0.6.13 for the on-arc observation, version 2.1.6.16 for the arc-interior
observation, and version 2.2.11.22 for the arc-background observation. We carried out further data
processing using HEAsoft\footnote{http://heasarc.gsfc.nasa.gov/lheasoft}
version~6.6 and CIAO\footnote{http://cxc.harvard.edu/ciao} version 3.4, following
guidelines available from the \suzaku\ Guest Observer Facility at
GSFC\footnote{http://suzaku.gsfc.nasa.gov/docs/suzaku/analysis/abc/abc.html}. We used
the set of calibration database (CALDB) files for the XIS released on 2009 February 03, and the CALDB
files for the X-ray Telescope (XRT) released on 2008 July 09. Throughout this paper we used only the
data from the back-illuminated XIS1 chip, as it is more sensitive at lower energies than the
front-illuminated chips.

\begin{deluxetable*}{lcccccc}
\tabletypesize{\footnotesize}
\tablewidth{0pt}
\tablecaption{Details of Our \suzaku\ Observations\label{tab:Observations}}
\tablehead{
			& \colhead{Observation}	& \colhead{$l$}		& \colhead{$b$}		& \colhead{Start time}	& \colhead{End time}	& \colhead{Usable exposure} 	\\
\colhead{Observation}	& \colhead{ID}		& \colhead{(deg)}	& \colhead{(deg)}	& \colhead{(UT)}	& \colhead{(UT)}	& \colhead{(ks)}
}
\startdata
Arc interior		& 502070010		& 253.29		& $-62.74$		& 2008-01-15 19:09:14	& 2008-01-17 18:20:14	& 74.3				\\
On arc			& 502071010		& 247.81		& $-64.51$		& 2007-06-05 07:29:21	& 2007-06-07 03:35:19	& 72.6				\\
Arc background		& 503104010		& 240.49		& $-66.01$		& 2008-12-30 06:07:54	& 2009-01-04 08:19:23	& 81.4				\\
\enddata
\end{deluxetable*}

\begin{figure*}
\centering
\begin{tabular}{ccc}
\includegraphics[width=0.3\linewidth,bb=72 217 470 575,clip=true]{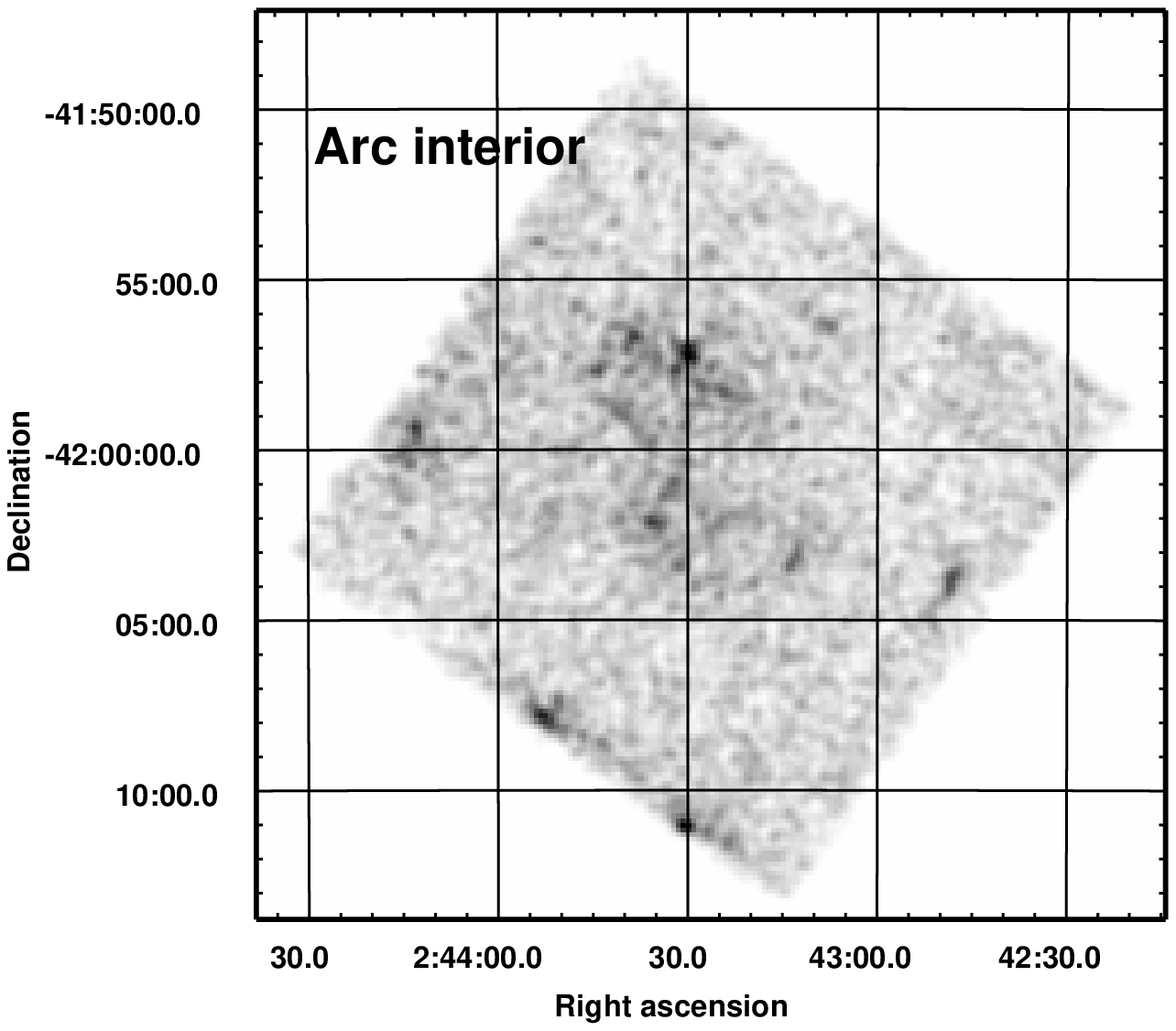} &
\includegraphics[width=0.3\linewidth,bb=72 217 470 575,clip=true]{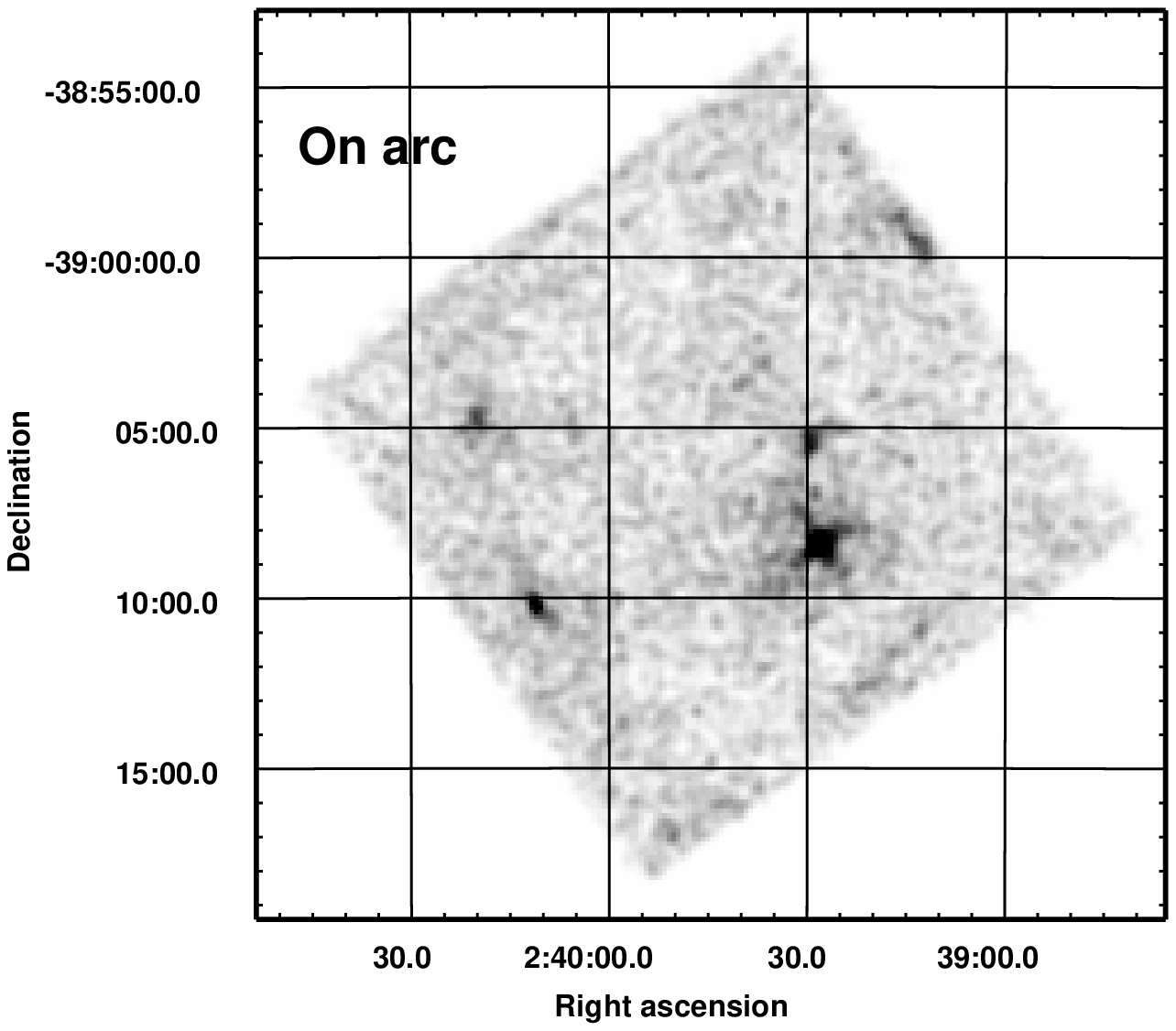} &
\includegraphics[width=0.3\linewidth,bb=72 217 470 575,clip=true]{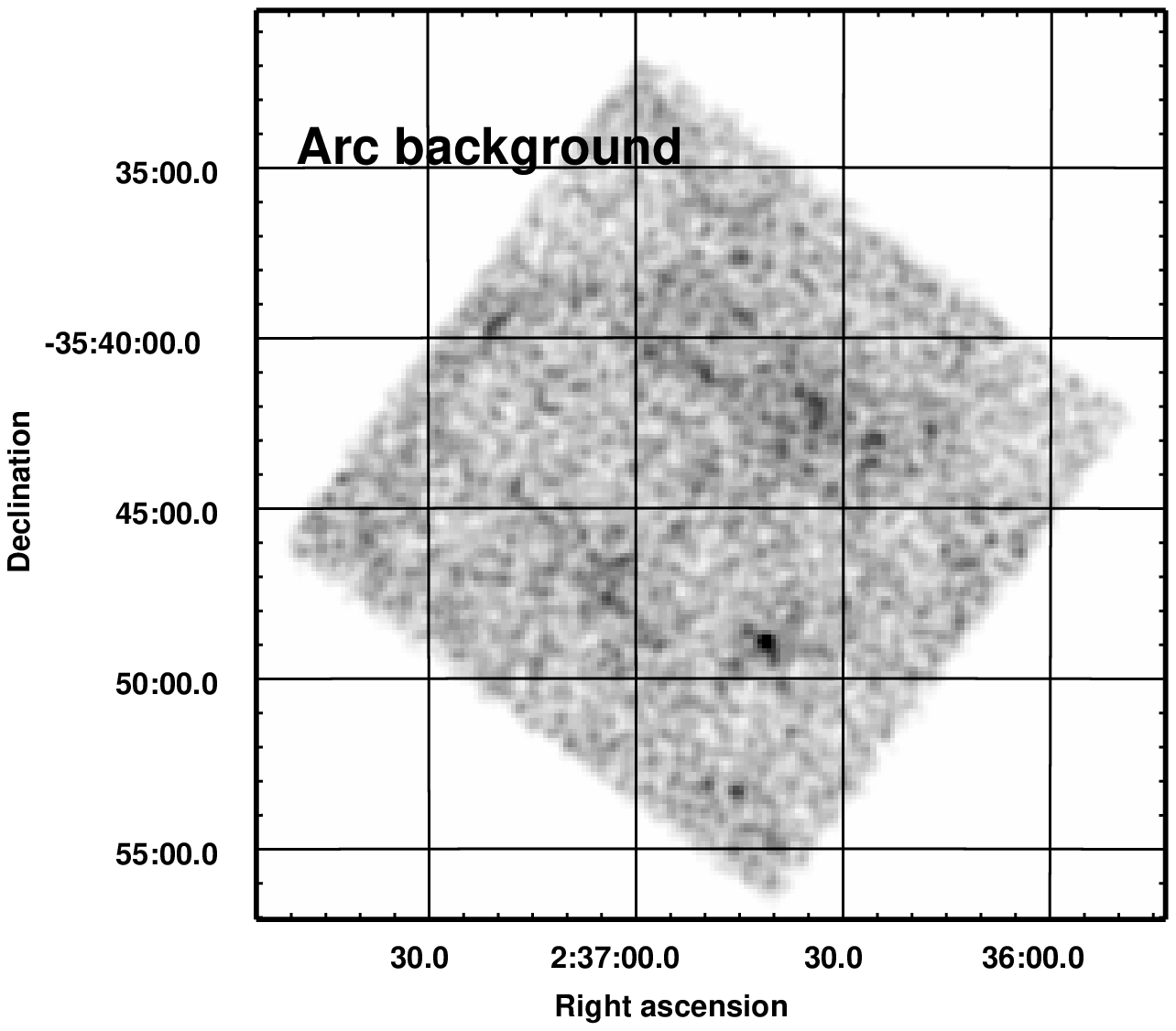}
\vspace{5mm}
\\
\includegraphics[width=0.29\linewidth,bb=86 217 526 575,clip=false]{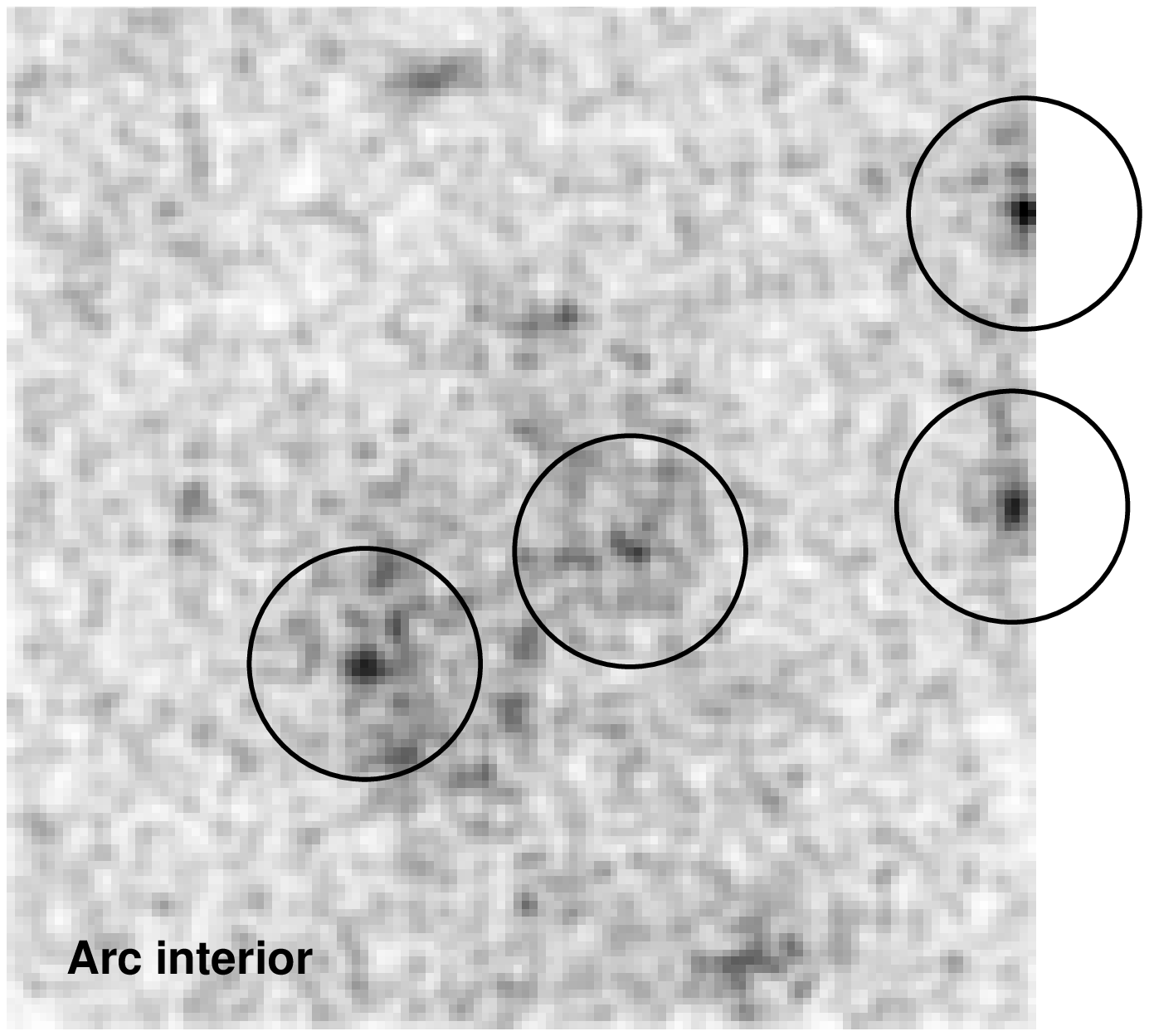} &
\includegraphics[width=0.29\linewidth,bb=86 217 526 575,clip=false]{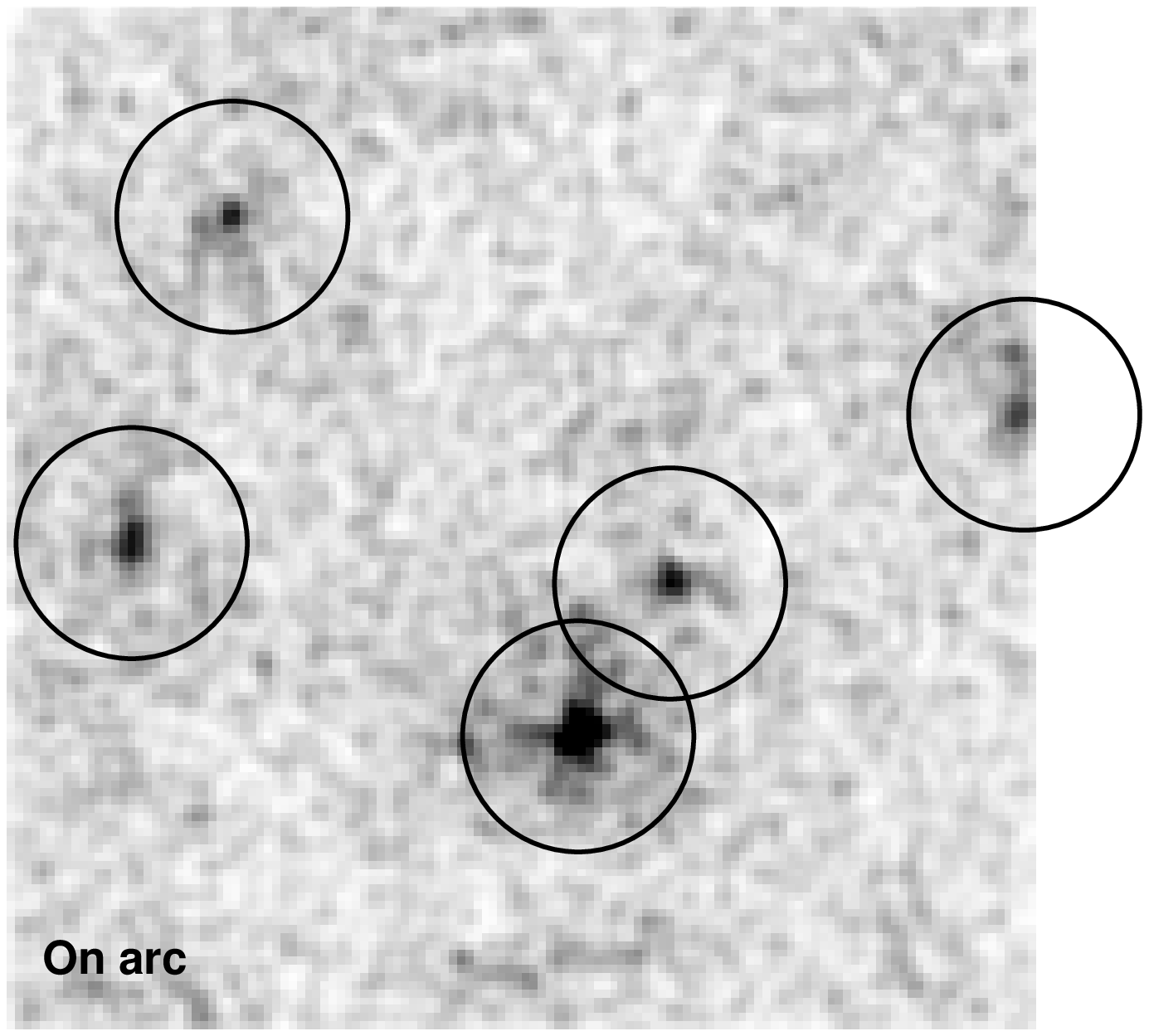} &
\includegraphics[width=0.29\linewidth,bb=86 217 526 575,clip=false]{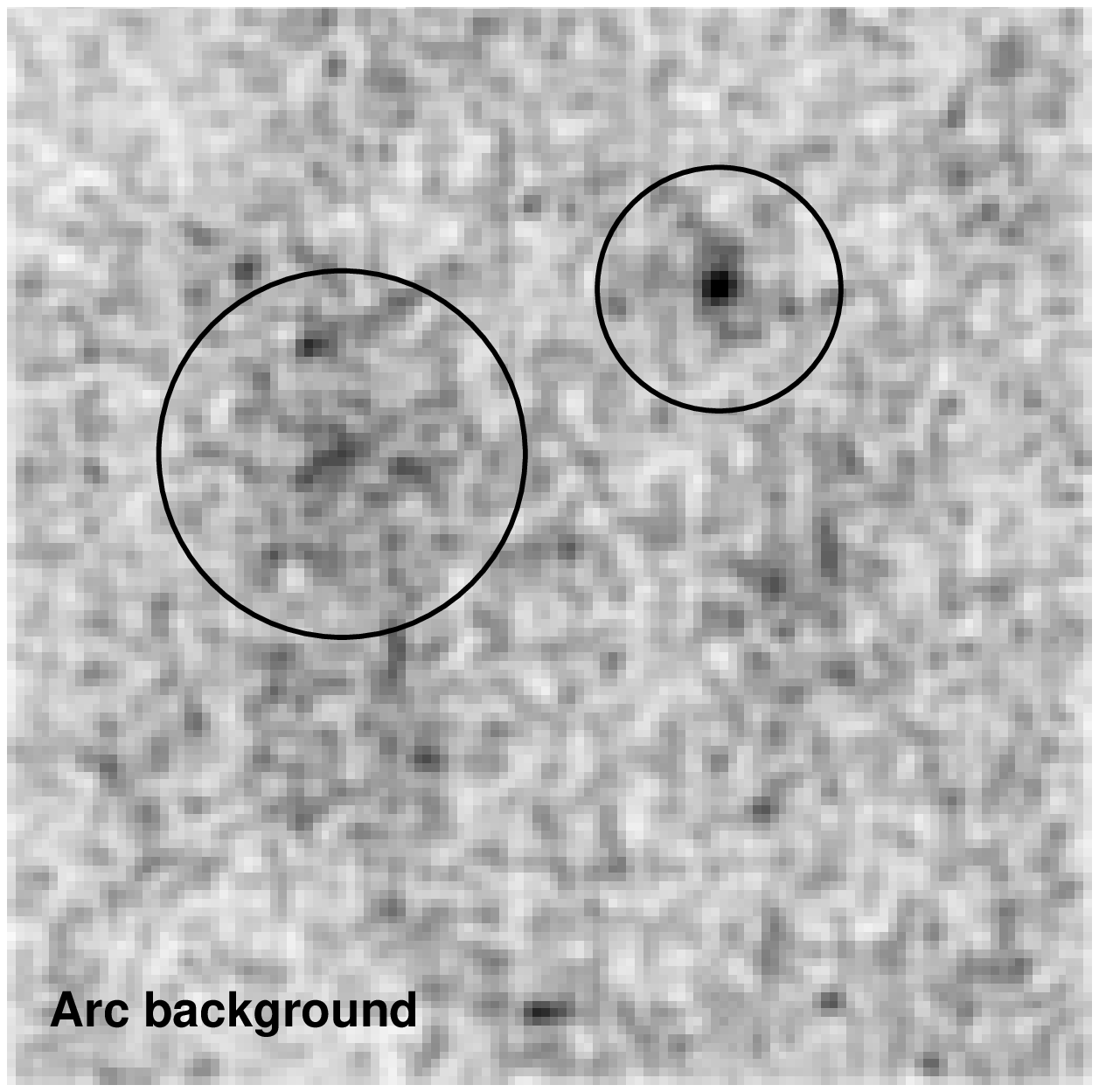}
\\
\end{tabular}
\caption{Cleaned and smoothed 0.3--5.0~\kev\ XIS1 images from our \suzaku\ arc observations.
The upper row of images shows the data in sky coordinates. The lower row
of images shows the data in detector coordinates. The black circles outline the regions that were excluded from the
analysis -- these circles were positioned by eye around sources that might contaminate the spectra of the diffuse
emission.\label{fig:XIS1images}}
\end{figure*}

We first used the \texttt{xispi} tool to update the XIS gain calibration. We then cleaned and filtered the data.  We
selected events with grades 0, 2, 3, 4, and 6, and used \texttt{cleansis} to remove flickering pixels.  We excluded the
times that \suzaku\ passed through the South Atlantic Anomaly (SAA), times up to 436~s after passage through the SAA,
times when \suzaku's line of sight was elevated less than 10\degr\ above the Earth's limb and/or was less than 20\degr\
from the bright-Earth terminator, and times when the cut-off rigidity (COR) was less than 8~GV. The thresholds used for
the elevation of \suzaku's line of sight above the Earth's limb and for the COR are higher than the defaults (which are
5\degr\ and 6~GV, respectively).  The higher elevation threshold reduces contamination from the scattering of
solar X-rays off the Earth's atmosphere, while the higher COR threshold reduces the particle background. We combined the
data taken in the 3$\times$3 and 5$\times$5 observation modes, and finally used the CIAO \texttt{analyze\_ltcrv.sl}
script to bin the 2.5--8.5~\kev\ data into 256-s time bins and remove times whose count-rates differ from the mean by
more than 3$\sigma$. The resulting cleaned XIS1 images are shown in Figure~\ref{fig:XIS1images}.

\subsection{Solar Wind Charge Exchange}

Observations of the diffuse soft X-ray emission from $\sim$million-degree Galactic gas may be contaminated
by geocoronal and heliospheric solar wind charge exchange (SWCX) emission
\citep{cravens00,cravens01,robertson03a,robertson03b,koutroumpa06,koutroumpa07}. Periods of enhanced SWCX
emission have been associated with periods of increased solar wind proton flux \citep{cravens01,snowden04,fujimoto07}.

\begin{figure}
\centering
\plotone{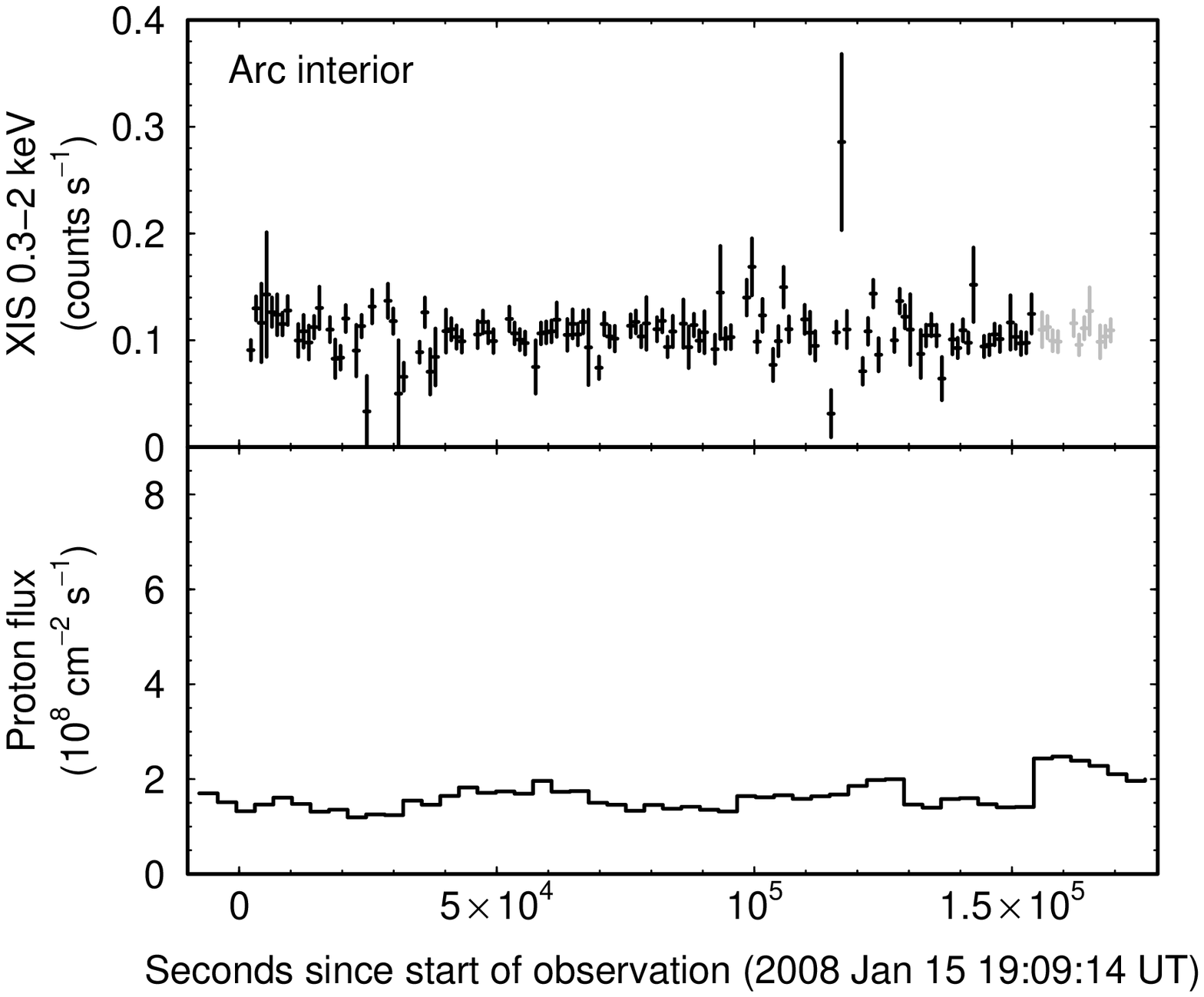} \\
\plotone{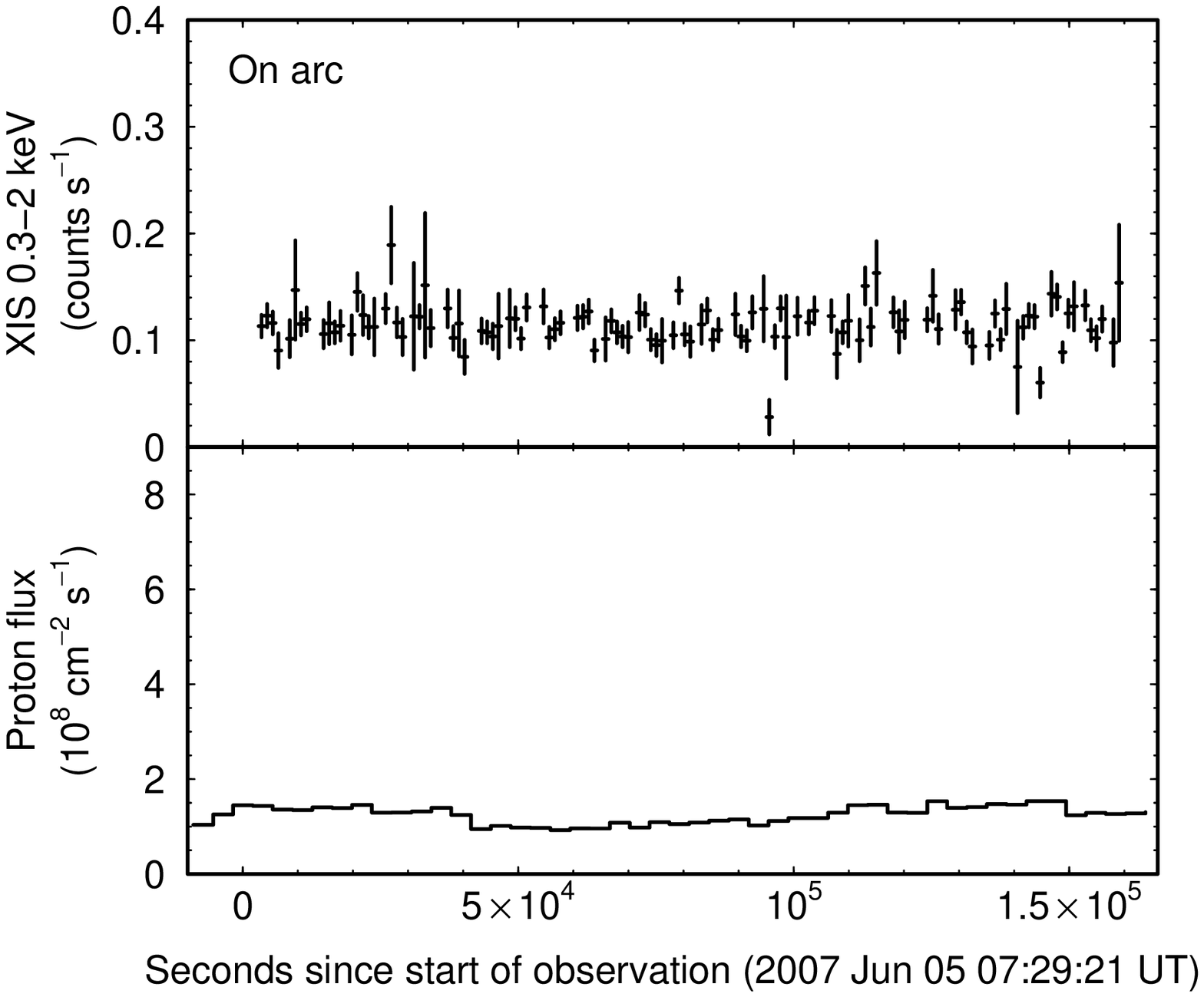} \\
\plotone{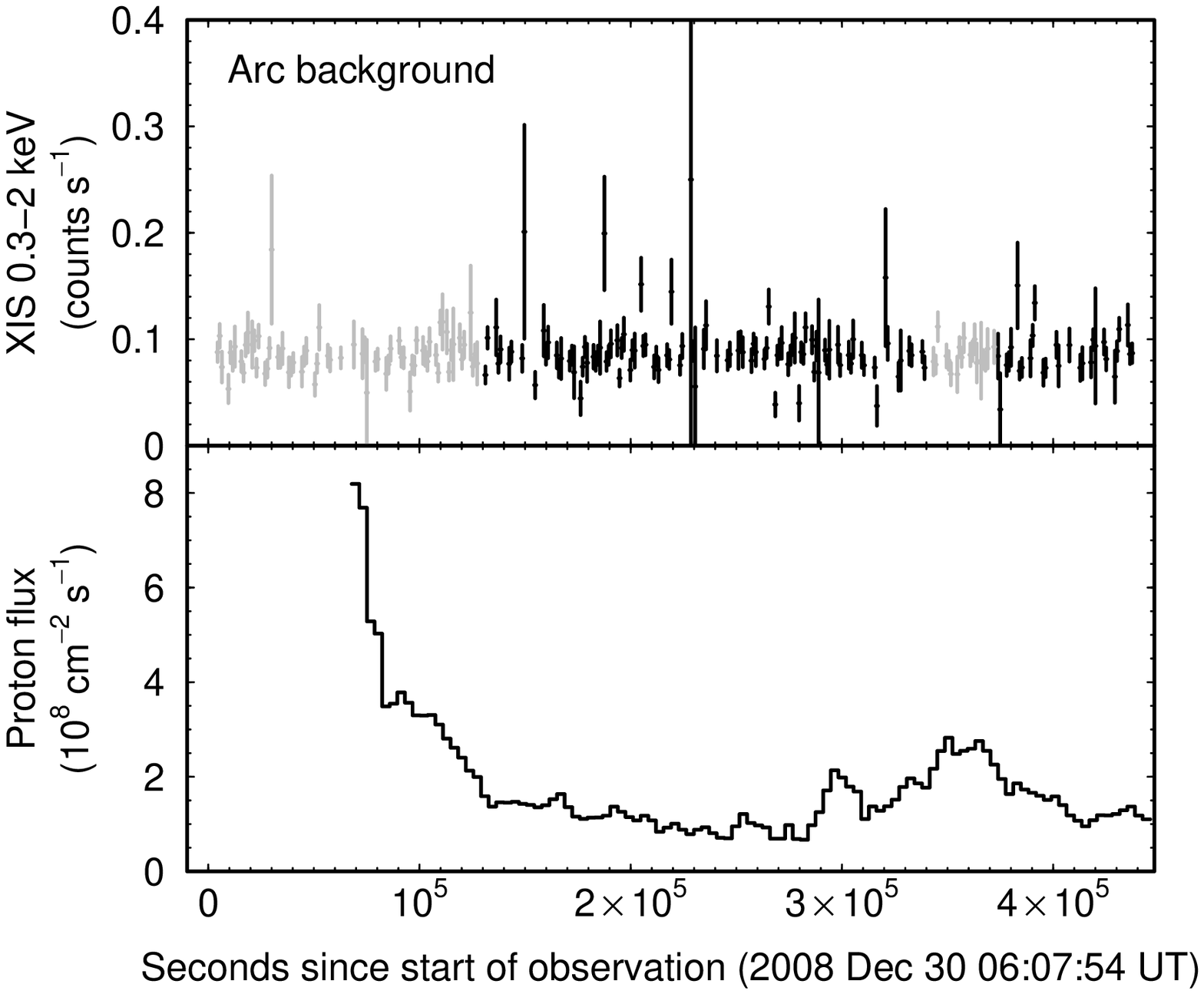}
\caption{XIS1 0.3--2.0~\kev\ lightcurves for each of our three observations, plotted alongside the contemporaneous
solar wind proton flux. No solar wind flux data are available for the first $\sim$70~ks of the arc-background
observation. The gray parts of the lightcurves correspond to times when solar wind data are missing, or
when the proton flux exceeded $2 \times 10^8$~\pcmsq~\ps. These times were removed from the data.
\label{fig:Lightcurves}}
\end{figure}

Figure~\ref{fig:Lightcurves} compares the 0.3--2.0~\kev\ XIS1 lightcurves for each of our three
\suzaku\ observations with the contemporaneous solar wind proton flux obtained from
OMNIWeb\footnote{http://omniweb.gsfc.nasa.gov/}, which combines solar wind data from several
different satellites. In particular, the solar wind data for the arc-interior observation are from
the \textit{Advanced Composition Explorer} (\ace), those for the on-arc observation are from \ace\
and \wind, and those for the arc-background observation are from \wind.

In general the solar wind proton flux was fairly steady during our observations. The most notable
exception to this statement is the first part of the arc-background observation, when the proton
flux was greatly increased. However, there was no significant increase in the soft X-ray count-rate
at this time. Despite the steadiness of the X-ray count-rate, we decided to err on the side of
caution and removed times when the proton flux exceeded $2 \times 10^8$~\pcmsq~\ps, or when no
solar wind data were available. These times are indicated by the gray datapoints in the lightcurves
in Figure~\ref{fig:Lightcurves}. The exposure times in Table~\ref{tab:Observations} are those
that remain after this additional filtering.

This procedure should help minimize contamination from bright, time-varying geocoronal SWCX. However,
it should be noted that it is possible that heliospheric SWCX emission and some quiescent geocoronal SWCX
emission still remain in our spectra.

\subsection{Point Source Removal}

We have previously found that automated source detection software does not work well on \suzaku\ images
\citep{henley08a}, presumably because of the broad point spread function.  We therefore identified
individual sources which may contaminate our diffuse spectra by eye from the 0.3--5.0~\kev\ XIS1
images. Because the tool that we used to calculate non-X-ray background spectra (\texttt{xisnxbgen}
version 2008-03-08) does not work correctly in sky coordinates, we worked in detector coordinates
when identifying and removing individual sources from our XIS1 images. We used circles of radius
115~pixels to exclude the events from around each identified source, except for one bright patch in
the arc-background observation, for which we used a radius of 173~pixels. These radii correspond to
$\approx$2\arcmin\ and $\approx$3\arcmin\ on the sky. The positions of the excluded sources are
shown in the lower row of Figure~\ref{fig:XIS1images}.

\subsection{Spectral Extraction, Non-X-ray Background, and Response Files}

We extracted spectra from the full XIS1 field of view, excluding the aforementioned sources and times, and binned the spectra
so there were at least 25 counts per bin. We used \texttt{xisnxbgen} to calculate non-X-ray background spectra from the
database of \suzaku\ night-Earth observations. For each observation we used the same parts of the XIS1 detector to
extract the source spectrum and to calculate the non-X-ray background spectrum.  We calculated redistribution matrix
files (RMFs) using \texttt{xisrmfgen}, and ancillary response files (ARFs) using \texttt{xissimarfgen}, which takes into
account the spatially varying contamination on the optical blocking filters (OBFs) of the XIS sensor
\citep{ishisaki07}. For the ARF calculations we assumed a uniform source of radius 20\arcmin.


\section{SPECTRAL ANALYSIS: EQUILIBRIUM MODEL}
\label{sec:CIEModels}

In this section we use a simple CIE model to look for variations in the halo emission in the
vicinity of the arc.  Our CIE model consists of unabsorbed emission from the LB and/or SWCX,
absorbed thermal emission from the Galactic halo, and an absorbed power-law for the extragalactic
background due to unresolved active galactic nuclei (AGN). The model is described in more detail in
\S\ref{subsec:ModelDescription}, and the results are presented in \S\ref{subsec:Results}.
Although this analysis does not directly address the question of whether or not the arc is the edge
of an extraplanar SNR, it establishes parts of the model that we use in our subsequent
analysis using SNR models (\S\ref{sec:SNRModels}), and it provides a benchmark for comparison with SNR models.

\subsection{Model Description}
\label{subsec:ModelDescription}

\subsubsection{The Foreground Component}
\label{subsubsec:ForegroundComponent}

When fitting a multicomponent model of the SXRB to a spectrum from a single direction, there can be
a degeneracy between the foreground (LB/SWCX) and background (halo + extragalactic) emission
components. We therefore used
\citepossessive{snowden00} catalog of shadows to fix the normalization of the foreground emission
in the vicinity of the arc. This catalog consists of 367 shadows in the 1/4~\kev\ RASS maps, five of
which are in the vicinity of the arc: S2425M653, S2555M621, S2560M560, S2646M675, S2658M582. For
each shadow, \citet{snowden00} decomposed the \rosat\ R1 and R2 count-rates into foreground and halo
count-rates (with associated errors). In the following test, we found that the foreground
count-rates for three of these five shadows could usefully constrain the foreground emission model.

We assumed that the foreground emission could be characterized with a $1T$ CIE plasma model
with $T \sim 10^6$~K, as such a model provides a reasonable fit to the \rosat\ data
\citep{snowden98,snowden00,kuntz00}, despite the fact that SWCX emission produces a
different spectrum from thermal plasma emission. At first we fitted a $1T$
\raymondsmith\ model with \citet{anders89} abundances to the foreground R1 and R2 count-rates for
the five \citet{snowden00} shadows near the arc, we obtained a temperature of $(3.8^{+0.3}_{-1.5}) \times 10^6$~K.
This foreground temperature is inconsistent with those found by previous studies of the SXRB
($T \sim 1.0$--$1.3 \times 10^6$~K; \citealt{snowden98,snowden00,kuntz00,galeazzi07,henley08a}).
In addition, this foreground model can be
ruled out because it predicts an \OVIII\ \Lyalpha\ intensity of $\approx$9~\lineunit\ (line units,
\LU), which far exceeds what is observed. A closer inspection of the shadows that we used showed
that the foreground R2/R1 ratios of two (S2646M675 and S2658M582) are inconsistent with $T \sim 1
\times 10^6$~K. When we eliminated these two shadows, we obtained a foreground emission model with
$T = 0.95 \times 10^6$~K and emission measure $\EMint = 0.0041$~\emismeas.  We used this foreground
model in our subsequent \suzaku\ analysis.

\subsubsection{The Halo and Extragalactic Components}

Although \citet{yao07a} and \citet{lei09} have shown that isothermal and two-temperature halo models
are inadequate for explaining all the available X-ray and UV emission and absorption data, these
models still serve a useful purpose in characterizing the X-ray portion of the emission spectra. Our
halo model consisted of a single CIE plasma component ($1T$ model), which was adequate to characterize
our data. The temperature and emission measure of the halo component were free parameters in the fitting.

We modeled the extragalactic background as an absorbed power-law. Its photon index was fixed at 1.46
\citep{chen97}, but its normalization was a free parameter.
So that we could model the attenuation of the halo and extragalactic components, we obtained
hydrogen columns densities from the Leiden-Argentine-Bonn (LAB) Survey of Galactic \HI\
\citep{kalberla05} using the HEAsoft \texttt{nh} tool. The hydrogen column densities for our three
observing directions are $\NH = 1.9 \times 10^{20}~\pcmsq$ (arc interior), $1.6 \times 10^{20}~\pcmsq$
(on arc), and $3.6 \times 10^{20}~\pcmsq$ (arc background).

\subsubsection{Additional Details}
\label{subsubsec:AdditionalDetails}

In order to better constrain the halo component at lower energies, we included R1 and R2 (1/4~\kev)
data from the RASS \citep{snowden97}.  These data were extracted from circles of radius 1\degr\
centered on each of our \suzaku\ pointing directions using the \texttt{sxrbg} tool available from
HEASARC\footnote{http://heasarc.gsfc.nasa.gov/Tools/xraybg\_help.html\#command}.  During the course
of our analysis, we discovered a discrepancy between our \suzaku\ spectra and the \rosat\ R45
(3/4~\kev) count-rates -- our models significantly underpredict the observed R45 count-rates.  We
will discuss this discrepancy in \S\ref{subsec:SuzakuR45DiscrepancyDiscussion}.  We
decided not to include the R45 data in our spectral analysis. We also did not include the R67 (1.5~\kev) data, as
this band is dominated by the extragalactic background, and it does not help constrain the Galactic
thermal emission.

We used XSPEC\footnote{http://heasarc.gsfc.nasa.gov/docs/xanadu/xspec/xspec11} version~11.3.2
\citep{arnaud96} to carry out the spectral analysis. For the thermal plasma components, we used the Astrophysical Plasma
Emission Code (APEC) version~1.3.1 \citep{smith01a} for the \suzaku\ spectra and the \raymondsmith\ code for the
\rosat\ R12 data. We used the \citeauthor{raymond77} code for the \rosat\ R12 data because APEC
is inaccurate in that band, due to a lack of data on transitions from L-shell ions of Ne, Mg, Al,
Si, S, Ar, and Ca.\footnote{http://cxc.harvard.edu/atomdb/issues\_caveats.html} The parameters of
each \citeauthor{raymond77} component in the \rosat\ model were tied to the parameters of the
corresponding APEC component in the \suzaku\ model (see \citealt{henley08a}). For the absorption, we
used the XSPEC \texttt{phabs} model, which uses cross-sections from \citet{balucinska92}, with an
updated He cross-section from \citet{yan98}. Throughout we used \citet{anders89} abundances.

We fitted the model to the 0.3--5.5~\kev\ \suzaku\ + \rosat\ R12 spectra, with the temperature and
emission measure of the halo components and the normalization of the extragalactic background as
free parameters. The low-energy cut-off for the \suzaku\ spectra was chosen because the XIS1
calibration is uncertain below 0.3~\kev. Although we did not expect much Galactic thermal emission
above $\sim$1~\kev, we included data up to 5.5~\kev\ in order to constrain the extragalactic
background. The high-energy cut-off was chosen to avoid the 5.9-\kev\ Mn \Kalpha\ line from the
radioactive $^{55}$Fe calibration source.

\subsection{Results for CIE Models}
\label{subsec:Results}

\begin{figure}
\centering
\plotone{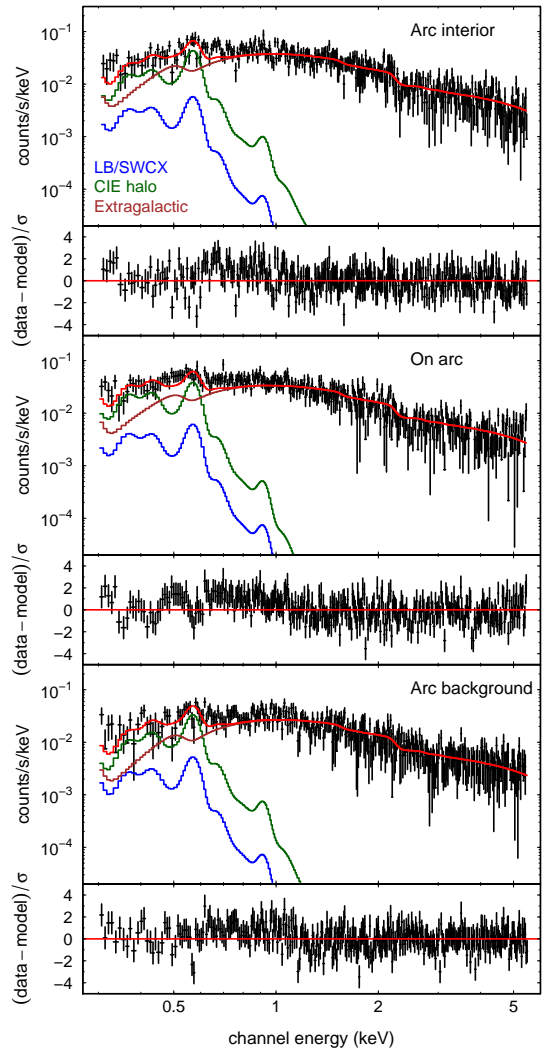}
\caption{Our observed arc-interior, on-arc, and arc-background \suzaku\ XIS1 spectra, plotted with
the best-fitting CIE models obtained by fitting to the \suzaku\ + R12 data for each direction individually.
\label{fig:Spectra+Model}}
\end{figure}

We fitted our LB/SWCX + $1T$ halo + extragalactic background model to each of our \suzaku\ + R12
spectra individually.  The results are shown in Table~\ref{tab:FitResults} and in
Figure~\ref{fig:Spectra+Model}. Generally, the fits are reasonably good. The agreement between the
model and the data is also good in the R12 band (not shown). The temperature of the halo is similar
in all three directions ($\sim$0.9--$1.1 \times 10^6$~K). However, the halo emission measure is
considerably larger in the on-arc direction -- this is not surprising, given that the arc is
brighter than its surroundings.

\begin{deluxetable*}{lccr@{/}l@{\,=\,}l}
\tabletypesize{\footnotesize}
\tablewidth{0pt}
\tablecaption{Spectral Fit Results\label{tab:FitResults}}
\tablehead{
			& \colhead{Halo $T$}				& \colhead{Halo E.M.\tablenotemark{a}}		& \multicolumn{3}{c}{}				\\
\colhead{Observation}	& \colhead{($10^6$ K)}				& \colhead{($10^{-3}$ \emismeas)}		& \multicolumn{3}{c}{$\chisq/\mbox{dof}$}	\\
}
\startdata
Arc interior		& $1.08 \pm 0.02$				& $17.8 \pm 0.6$				&  400.67 &  346 & 1.16		\\
On arc			& $0.95 \pm 0.02$				& $30.0^{+1.0}_{-1.2}$				&  394.71 &  322 & 1.23		\\
Arc background		& $1.03 \pm 0.02$				& $23.3 \pm 1.8$				&  371.31 &  339 & 1.10		\\
\enddata
\tablenotetext{a}{Emission measure $\mbox{E.M.} = \EMint$}
\end{deluxetable*}

The fact that the halo emission measure is larger in the on-arc direction than the arc-background
direction supports statement in the Introduction that the arc's shape is not due to absorption.
If its shape were due to absorption, we would expect the halo to have similar intrinsic intensities
in these two directions -- this is not what is observed. The halo is slightly cooler in the
on-arc direction compared with other two directions, although the difference in temperature is
rather small.

Although not relevant to the subsequent discussion, for completeness we note that the normalizations of the
extragalactic background in the three directions were 10.6 (arc interior), 9.8 (on arc),
and 8.1 (arc background) \pownorm\ at 1~\kev.


\section{SPECTRAL ANALYSIS: SUPERNOVA REMNANT MODELS}
\label{sec:SNRModels}

The above-described analysis shows that there is intrinsic variation in the halo's spectrum on
and around the arc. However, by itself the above analysis has not helped us address the question of
whether or not the arc is the edge of an extraplanar SNR. In this section, we test this hypothesis
by fitting spectral models derived from hydrodynamical simulations of extraplanar SNRs \citep{shelton06}
to our \suzaku\ + R12 spectra. We look at SNR models with a range of ambient densities (corresponding
to different heights above the disk), ambient magnetic fields, and ages. We will assess these models
on the basis of the spectrum, brightness and gross morphology of the arc.

The hydrodynamical simulations from which our spectral models are derived are described in
\S\ref{subsec:SNRHydroModels}.  Our spectral model is described in
\S\ref{subsec:SNRModelDescription}. In \S\ref{subsec:SNRResults} we present the results of the
spectral fitting, and we identify the SNR models that are in best agreement with the
observations. Finally, in \S\ref{subsec:SNRProfiles} we compare the radial surface brightness
profile predicted by one of our best-fitting models with the observed \rosat\ R12 and R45 profiles.

\subsection{Hydrodynamical Simulations of Extraplanar Supernova Remnants}
\label{subsec:SNRHydroModels}

\citet{shelton06} carried out 1-D hydrodynamical simulations of extraplanar SNRs in 7 different ambient densities
($n_0 = 0.5$, 0.2, 0.1, 0.05, 0.02, 0.01, 0.005~\pcc), corresponding to heights, $z$, above the
midplane ranging from 76 to 1800~\pc\ (using the interstellar density model from \citealt{ferriere98a}).
For each ambient density, she carried out simulations with
$(E_0 / 10^{51}~\erg, \Beff / \microgauss) = (0.5, 2.5)$ (which we call model type B), $(0.5, 5.0)$
(model type C), and $(1.0, 5.0)$ (model type D), where $E_0$ is the explosion energy and \Beff\ is
the effective magnetic field, which produces a nonthermal pressure in addition to the ambient gas
pressure. We have added a fourth type of model, with $E_0 = 0.5 \times 10^{51}~\erg$ and $\Beff = 0$
(model type A). The model parameters are summarized in Table~\ref{tab:ModelParameters} (cf.
Table~1 in \citealt{shelton06}). Table~\ref{tab:ModelParameters} gives the conversion between $z$
and $n_0$, as well as model IDs that we will use below. The numerical part of the model ID indicates
the height of the SNR in \pc, while the letter (A--D) indicates $E_0$ and \Beff.
The models include thermal conduction, and the ionization evolution in the shocked gas is modeled self-consistently.

\begin{deluxetable}{ccccc}
\tabletypesize{\footnotesize}
\tablewidth{0pt}
\tablecaption{SNR Model Parameters\label{tab:ModelParameters}}
\tablehead{
\colhead{$z$}		& \colhead{$n_0$}	& \colhead{$E_0$}		& \colhead{\Beff}		& \colhead{Model}	\\
\colhead{(\pc)}		& \colhead{(\pcc)}	& \colhead{($10^{51}~\erg$)}	& \colhead{(\microgauss)}	& \colhead{ID}		\\
}
\startdata
76			& 0.5			& 0.5				& 0				& 76A			\\
			& 0.5			& 0.5				& 2.5				& 76B			\\
			& 0.5			& 0.5				& 5.0				& 76C			\\
			& 0.5			& 1.0				& 5.0				& 76D			\\
190			& 0.2			& 0.5				& 0				& 190A			\\
			& 0.2			& 0.5				& 2.5				& 190B			\\
			& 0.2			& 0.5				& 5.0				& 190C			\\
			& 0.2			& 1.0				& 5.0				& 190D			\\
310			& 0.1			& 0.5				& 0				& 310A			\\
			& 0.1			& 0.5				& 2.5				& 310B			\\
			& 0.1			& 0.5				& 5.0				& 310C			\\
			& 0.1			& 1.0				& 5.0				& 310D			\\
480			& 0.05			& 0.5				& 0				& 480A			\\
			& 0.05			& 0.5				& 2.5				& 480B			\\
			& 0.05			& 0.5				& 5.0				& 480C			\\
			& 0.05			& 1.0				& 5.0				& 480D			\\
850			& 0.02			& 0.5				& 0				& 850A			\\
			& 0.02			& 0.5				& 2.5				& 850B			\\
			& 0.02			& 0.5				& 5.0				& 850C			\\
			& 0.02			& 1.0				& 5.0				& 850D			\\
1300			& 0.01			& 0.5				& 0				& 1300A			\\
			& 0.01			& 0.5				& 2.5				& 1300B			\\
			& 0.01			& 0.5				& 5.0				& 1300C			\\
			& 0.01			& 1.0				& 5.0				& 1300D			\\
1800			& 0.005			& 0.5				& 0				& 1800A			\\
			& 0.005			& 0.5				& 2.5				& 1800B			\\
			& 0.005			& 0.5				& 5.0				& 1800C			\\
			& 0.005			& 1.0				& 5.0				& 1800D			\\
\enddata
\tablecomments{$\Beff = 2.5$ and 5.0~\microgauss\ correspond to non-thermal pressures of 1800 and 7200~\presalt, respectively.}
\end{deluxetable}

\begin{figure}
\centering
\plotone{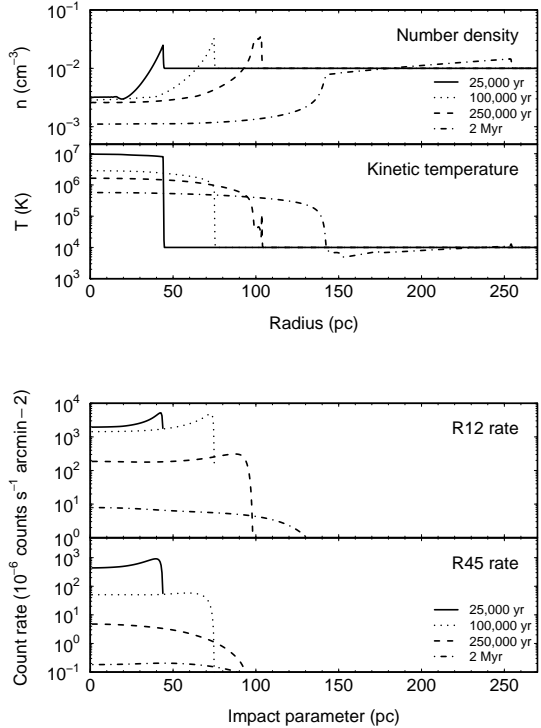}
\caption{The evolution of the density and temperature as a function of radius, and the predicted
\rosat\ R12 and R45 count-rates as a function of impact parameter, for SNR model 1300B
(see text for details).
\label{fig:SNRModelProfiles}}
\end{figure}

Figure~\ref{fig:SNRModelProfiles} shows how one such SNR evolves in time. This model is model
1300B in Table~\ref{tab:ModelParameters}, and remnant~A in \citet{shelton99}; see that paper for
more details. At early times (solid line, $t = 25,000~\yr$), the explosion creates a hot
cavity or bubble, bounded by the shock. The bubble produces copious soft X-rays -- these X-rays come
mainly from the hot dense region behind the shock, and the remnant is edge-brightened. As the bubble
expands (dotted line, $t = 100,000~\yr$), it cools by adiabatic expansion. Nevertheless,
the R12 emission remains bright and edge-brightened, although the R45 emission drops
considerably. Between $t = 100,000$ and 250,000~yr (dashed line), the dense rim of the
remnant undergoes rapid radiative cooling, forming a cool, dense shell between the edge of the hot
bubble at $\approx$100~pc and the shock at $\approx$105~pc. The X-ray count-rate drops considerably
after this occurs, although in this case the R12 emission remains edge-brightened.  As time goes on,
the hot bubble continues to cool (both adiabatically and radiatively) and it gets progressively
fainter in X-rays, while the cool shell widens. Eventually, the X-ray emission ceases to be
edge-brightened (dot-dash line, $t = 2~\Myr$), before the remnant eventually fades away
altogether.

Note that the times quoted above are specific to the chosen model -- different model parameters
(particularly the ambient density) will result in changes on different timescales.  For example, a
model with $n_0 = 0.5~\pcc$ (corresponding to $z = 76~\pc$) will form a cool shell between $t =
25,000$ and 50,000~yr, and the R12 emission will cease to be edge-brightened by $t = 500,000~\yr$.
However, the general pattern of change described above applies to all the models.  As
described in the Introduction, \citet{shelton06} showed that the combined emission from an ensemble
of such SNRs of different ages and at different heights could explain a large fraction of the
observed 1/4~\kev\ halo emission at high Galactic latitudes. Here, we use the results of the SNR
simulations summarized in Table~\ref{tab:ModelParameters} to calculate model spectra which we
compare with our \suzaku\ + R12 spectra.

\subsection{Spectral Model Description}
\label{subsec:SNRModelDescription}

Each simulation dataset consists of hydrodynamical data and ion populations as a function of radius
for up to 26 epochs, ranging from $t = 2.5 \times 10^3$ to $2 \times 10^7$~yr (the range of ages
covered depends on the model).  These data were used to calculate X-ray spectra. As the \suzaku\
field of view is much smaller than the size of the arc, we did not calculate spectra for the whole
model remnant, but instead calculated projected spectra along various sightlines through the
remnant. For this purpose, we used software developed by \citet{shelton99}, which uses the
\citet{raymond77} spectral code (updated by J.~C. Raymond \& B.~W. Smith, 1993, private
communication with R.~J. Edgar) with abundances from \citet{anders89}. The spectral calculations
take into account the (possibly non-equilibrium) ionization fractions output by the hydrodynamical
simulations.

For each epoch of each SNR model, we calculated projected spectra for sight lines through the SNR at
various impact parameters (each was measured from the projected center of the remnant). We
normalized the model impact parameters such that the normalized value was 0 for a sightline
through the center and 1 for a sightline through the part of the rim at which the model R12 count-rate
was greatest.  For example, for the model shown in Figure~\ref{fig:SNRModelProfiles}, the model
impact parameters were normalized by dividing by 42.2~\pc\ at $t = 25,000~\yr$, and by 72.7~\pc\ at
$t = 100,000~\yr$. We then used these sets of projected spectra as a function of this normalized
impact parameter to create XSPEC table
models\footnote{http://xspec.gsfc.nasa.gov/docs/xanadu/xspec/xspec11/manual/node61.html}, one for
each epoch of each SNR model.

By eye, we estimated the center of the hypothesized bubble to be at
$(l,b) = (256.014\degr,-61.575\degr)$ (assuming that the bubble is a circle), while the brightest
part of the arc (in the R12 band) is at $(l,b) = (247.90\degr,-65.09\degr)$. Therefore, the impact
parameters of our on-arc and arc-interior sightlines, normalized to the impact parameter at which
the R12 count-rate is greatest, are 0.936 and 0.341, respectively.  We used projected SNR spectra
calculated for these normalized impact parameters in our spectral analysis.

As in \S\ref{sec:CIEModels}, we created a multicomponent emission model consisting of LB/SWCX, halo,
and extragalactic components. However, instead of using a $1T$ CIE model for the halo, we used a SNR
+ $1T$ CIE model.  The normalization of the brightness of the SNR component was a free parameter, i.e., the brightness
of the model SNR emission could be scaled up or down to best match the observed spectrum. If a given
SNR model accurately predicts the arc emission, its normalization should be 1. If the spectrum
predicted by the SNR component of our multicomponent model is too bright, the SNR model
normalization will be less than 1. Similarly, if the model SNR spectrum is too faint, the SNR model
normalization will be greater than 1. The normalization of the SNR component was constrained to be
the same for the on-arc and arc-interior directions, and was set to zero for the arc-background
direction (i.e., we assumed no SNR emission was contributing to the arc-background spectrum).
The CIE portion of the halo component was constrained to be the same in all three directions.
As described below, the CIE component is needed to provide adequate fits to the data. This CIE
component represents uniform hot halo gas that is not due a young SNR. This hot gas is of
unknown origin, and could be due to old SNRs, outflows from the disk, or infalling extragalactic
material.

For the LB/SWCX emission, we used the model described in \S\ref{subsubsec:ForegroundComponent}. For the
extragalactic background, we again used a power-law with a photon index of 1.46, the normalization
of which was independent for each of the three observation directions.

For each epoch of each SNR model, we fitted this LB/SWCX + CIE~halo + SNR + extragalactic model to
our three \suzaku\ + R12 spectra simultaneously. The free parameters during the fitting were the
normalization of the SNR component, the temperature and emission measure of the CIE halo component, and
the normalizations of the extragalactic background.

\subsection{Results for SNR Models}
\label{subsec:SNRResults}

Here we evaluate our SNR models on the basis of their spectrum (which models give the lowest
\chisq?), their brightness (which SNR components' normalizations are closest to 1?), and their
gross morphology (which models predict an edge-brightened remnant whose radius is close to
that of the arc?).

\begin{figure}
\centering
\plotone{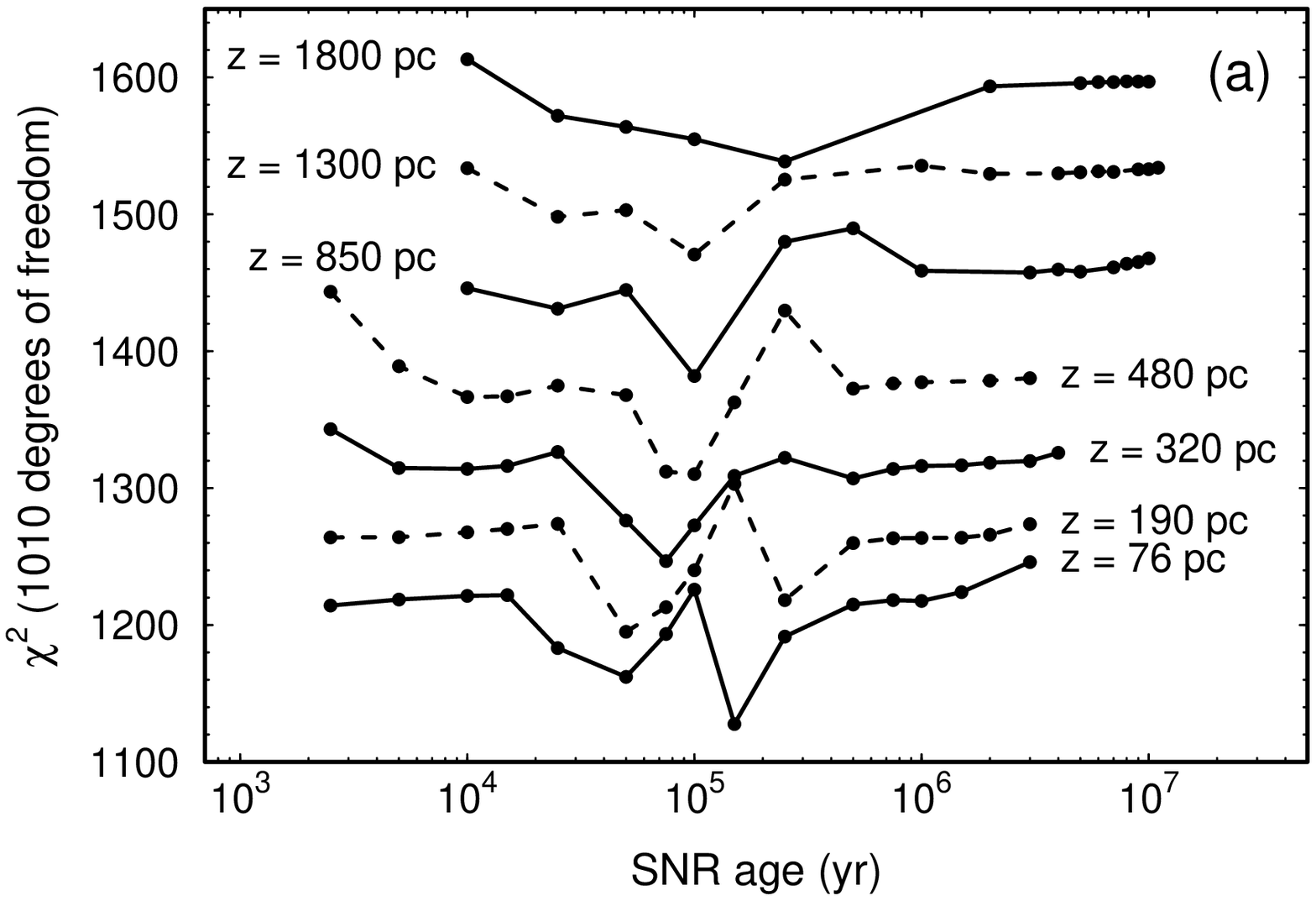}
\plotone{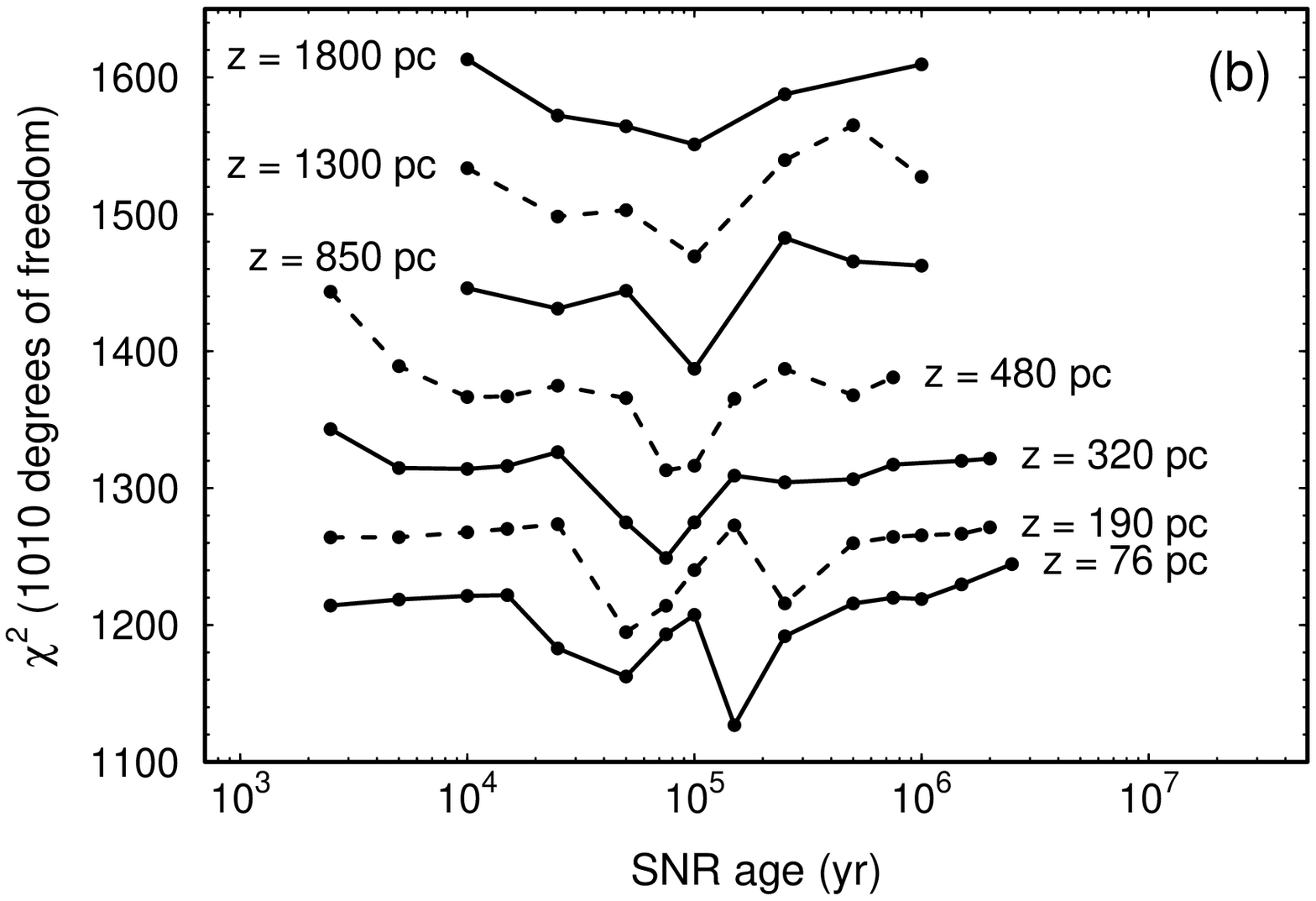}
\caption{Variation of \chisq\ with SNR age for SNR models at different heights, $z$. These results are from
fitting the LB/SWCX + CIE~halo + SNR + extragalactic model described in \S\ref{subsec:SNRModelDescription} to our three
\suzaku\ + R12 spectra, using (a) SNR model type A ($E_0 = 0.5 \times 10^{51}~\erg$, $\Beff = 0$)
and (b) SNR model type B ($E_0 = 0.5 \times 10^{51}~\erg$, $\Beff = 2.5~\microgauss$).
For clarity, the curves for $z = 190$, 310, ..., 1800~\pc\ have been shifted upward by 50, 100,..., 300.
\label{fig:chisq-vs-age}}
\end{figure}

Simultaneously fitting our LB/SWCX + CIE~halo + SNR + extragalactic model to our three spectra yielded sets of best-fit
parameters as a function of SNR age for each of the 28 SNR models. Figure~\ref{fig:chisq-vs-age}
shows \chisq\ as a function of SNR age for a subset of these SNR models. The results plotted are
specifically for (a) model type A, with $E_0 = 0.5 \times 10^{51}~\erg$ and $\Beff = 0$,
and (b) model type B, with $E_0 = 0.5 \times 10^{51}~\erg$ and $\Beff = 2.5~\microgauss$,
but the other SNR models produce similar curves. The different curves correspond to SNe exploding at
a series of increasing heights (that is, in a series of decreasing ambient densities). Note that
not every model epoch yielded a valid fit -- at later epochs the model SNR is brightest in the
center, whereas the arc (if it is a SNR) is edge-brightened. As a result, not every epoch is shown
in the curves in Figure~\ref{fig:chisq-vs-age}. The CIE halo components from these fits have
temperatures of $\sim$0.8--$1.1 \times 10^6~\K$ and emission measures of $\sim$0.015--0.03~\emismeas.
The fits are considerably worse without this component. For example, without a halo component, the
arc-background R12 count-rate is significantly underpredicted ($602 \times 10^{-6}~\rassrate$,
against an observed count-rate of $(794 \pm 13) \times 10^{-6}~\rassrate$). Without this component,
the best-fitting model epochs give $\chi^2 \sim 1600$ for 1012 degrees of freedom, against
$\chi^2 \sim 1200$ for 1010 degrees of freedom with the CIE component.

For the lowest height ($z = 76~\pc$) there are two minima in the \chisq\ curves,
corresponding to SNR ages of 50,000 and 150,000~yr.  However, for $z \ge 320~\pc$, there is a clear
single minimum in the \chisq\ curves, corresponding to SNR ages of 75,000--100,000~yr. (The exception
is the model 1800A, for which the best-fitting age is 250,000~yr.)

\begin{figure}
\centering
\plotone{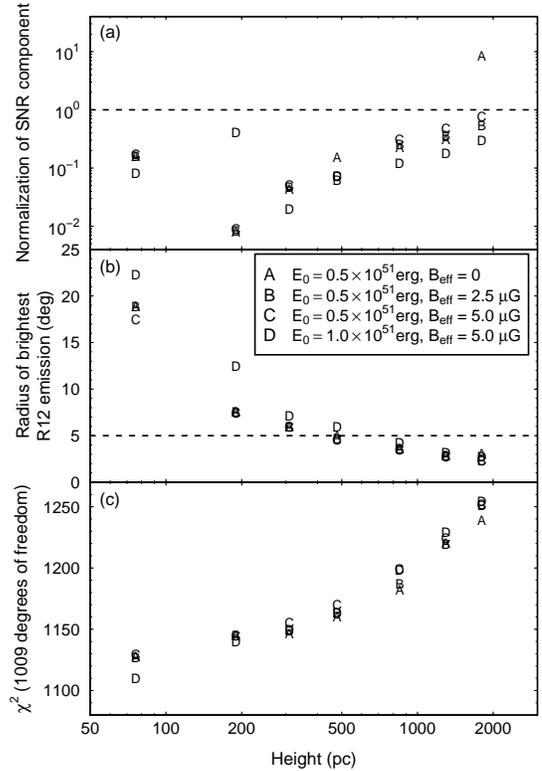}
\caption{The best-fitting (a) normalization of the SNR component, (b) radius of brightest
R12 emission, and (c) \chisq\ for each of the 28 SNR models that we
investigated. The results are plotted as a function of the height of the original SN explosion,
corresponding to different ambient densities (see Table~1 in \citealp{shelton06}). The letters
A--D denote models with different explosion energies, $E_0$, and effective ambient magnetic
fields, \Beff\ (see Table~\ref{tab:ModelParameters}).
The horizontal dashed lines denote a SNR normalization of 1, and a
radius of brightest R12 emission of 5\degr\ (see text for details). Note that the values of \chisq\
are for one fewer degree of freedom than in Figure~\ref{fig:chisq-vs-age}, because for each model
we have selected the epoch with the lowest \chisq.
\label{fig:Norm+Radius}}
\end{figure}

The above-described curves tell us which SNR age gives the best fit to the spectra for a given SNR
model. However, we would like to further discriminate among the models. To this end, we consider the
normalizations of the SNR components, the predicted radii at which the R12 emission is brightest, and
\chisq. Figure~\ref{fig:Norm+Radius}(a) shows the best-fitting normalization of the SNR component
for each of the 28 SNR models. As was noted above, if a SNR spectral model accurately represents the
arc emission, then its normalization should be 1 -- this is shown by a dashed line in the
plot. Figure~\ref{fig:Norm+Radius}(b) shows the predicted radii at which the R12 emission
peaks. These radii were calculated for each SNR model using the best-fitting ages. The
hydrodynamical simulations give these radii in parsecs. To convert these to angular radii, we
calculated the distances to the model SNRs using the nominal heights of the SNR models and assuming
a Galactic latitude of $-60$\degr. If the center of the hypothesized bubble is at $(l,b) =
(256.014\degr,-61.575\degr)$ (see above), the observed radius at which the R12 emission peaks is
$\approx$5\degr\ -- this is shown by a dashed line in the plot. Figure~\ref{fig:Norm+Radius}(c)
shows the best-fit \chisq\ for each model.

For $z = 76~\pc$, the best-fitting SNR models are $\sim$6--12 times too bright, and a factor of
$\sim$4 too large. For $z = 190$ and 310~\pc, the predicted radii are within a factor of 2 of the
observed value, but the best-fitting models are $\sim$20-120 times too bright. The exception
is model 190D, which is less than 3 times too bright. However, its predicted radius is 2.5 times
too large. For $z = 480~\pc$ the predicted radii are in very good agreement with the observed value,
but the best-fitting models are still an order of magnitude too bright.

The best agreement between the models and the observations is for $z \ge 850~\pc$. In terms of the
normalization of the SNR component, the models with $E_0 = 0.5 \times 10^{51}~\erg$ (model types A, B, and C) are
better than those with $E_0 = 1.0 \times10^{51}~\erg$ (model type D).  The best-fitting $E_0 = 0.5
\times 10^{51}~\erg$ models are less than a factor of 4.5 too bright, and the radii are within a
factor of 2.2 of the observed value (the exception is model 1800A, which is a factor of $\sim$10
too faint). Apart from this one model, all models for $z \ge 850~\pc$ give a best-fitting SNR age of
100,000~yr. However, we cannot easily discriminate among the models for $z \ge 850~\pc$ with $E_0 =
0.5 \times 10^{51}~\erg$, apart from ruling out model 1800A. For example, models 1800B and 1800C
have the best-fit SNR normalizations closest to 1, but these models also have larger values of 
\chisq\ than the models for 850 or 1300~\pc. Also, at a given height, all the models have
similar values of \chisq.

\begin{figure}
\centering
\plotone{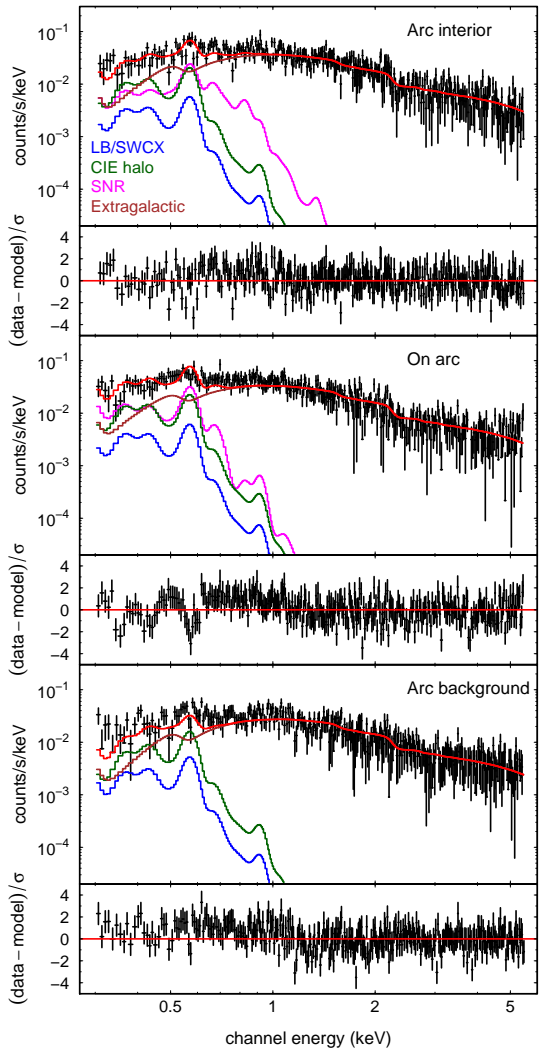}
\caption{The three \suzaku\ spectra with the best-fitting LB/SWCX + CIE~halo + SNR + extragalactic model obtained using
SNR model 1300B ($E_0 = 0.5 \times 10^{51}~\erg$, $\Beff = 2.5~\microgauss$). The individual model components
are also plotted.
\label{fig:Spectra+SNRModel}}
\end{figure}

Figure~\ref{fig:Spectra+SNRModel} shows our three \suzaku\ spectra along with the best-fitting spectral model
obtained from the SNR model 1300B ($E_0 = 0.5 \times 10^{51}~\erg$, $\Beff = 2.5~\microgauss$).
This model is in the middle of the range of acceptable heights, and in the middle of the
range of magnetic fields used.
The model generally does a reasonable job of fitting the spectra (including the R12 data, which are
not shown), although the $\sim$0.6--0.9~\kev\ flux is underpredicted in all three spectra (in the
arc-background spectrum, the flux is also underpredicted at lower energies). Over the whole \suzaku\
band, the SNR component is brighter in the on arc direction. However, above 0.6~\kev\ the SNR
component is brighter in the arc-interior direction, implying that the X-ray emission is harder
toward the center of the SNR than toward the edge.

\subsection{Comparison of the SNR Model with R12 and R45 Profiles across the Arc}
\label{subsec:SNRProfiles}

\begin{figure}
\plotone{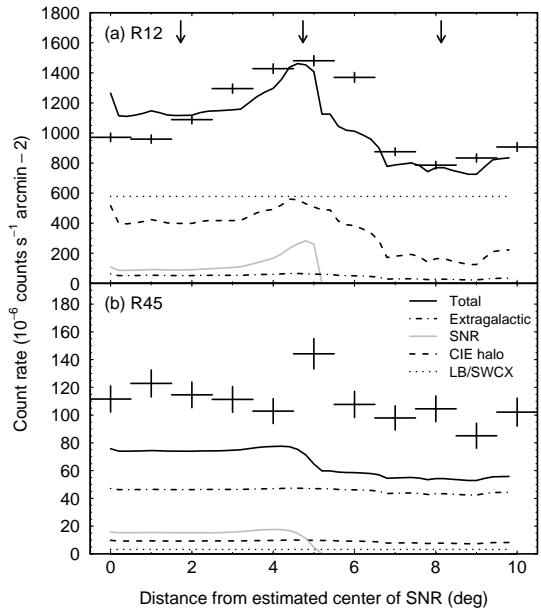}
\caption{Profiles of the observed (a) R12 and (b) R45 count-rates across the arc (crosses)
compared with the profile predicted by our best-fitting LB/SWCX + CIE~halo + SNR + extragalactic
model obtained using SNR model 1300B ($E_0 = 0.5 \times 10^{51}~\erg$, $\Beff = 2.5~\microgauss$)
(solid line).  The individual model components are also plotted. The horizontal axis shows
the distance from the estimated center of the hypothesized SNR at $l = 256.014\degr$, $b = -61.575\degr$.
The arrows show the positions of the three \suzaku\ observations (left to right: arc interior, on arc, arc background).
\label{fig:SNRROSATProfiles}}
\end{figure}

An additional comparison we can make between our model and the observations is to look at profiles
of the \rosat\ R12 and R45 count-rate across the arc. Such a comparison is made in
Figure~\ref{fig:SNRROSATProfiles}, which shows (a) R12 and (b) R45 profiles along
a great circle on the celestial sphere starting at the estimated center of the hypothesized SNR at
$l = 256.014\degr$, $b = -61.575\degr$, and passing over the arc, as near as possible to the three
\suzaku\ observations. The model profiles were calculated using the best-fit parameters obtained in
the previous section using SNR model 1300B ($E_0 = 0.5 \times 10^{51}~\erg$,
$\Beff = 2.5~\microgauss$). For each direction along the model profile, we used the column density
from \citet{kalberla05} calculated using the HEAsoft \texttt{nh} tool. Although we found some variation
in the normalization of the extragalactic background among our three \suzaku\ spectra (see \S\ref{subsec:Results}),
we do not know how the extragalactic background varies in the directions between our \suzaku\ observations.
Therefore, for the purposes of Figure~\ref{fig:SNRROSATProfiles} we assumed a constant value of 9~\pownorm\ at 1~keV.
The conclusions drawn from Figure~\ref{fig:SNRROSATProfiles} are unlikely to be affected by relaxing that assumption.

Figure~\ref{fig:SNRROSATProfiles}(a) shows that the variation in the model R12 intensity across the arc is due
both to the edge-brightened SNR, and to variations in the CIE halo intensity (which are due to variations in \NH,
as this model component does not vary intrinsically). The model overpredicts the R12 intensity near the center
of the hypothesized bubble, and underpredicts the width of the arc.

In the R45 band (Figure~\ref{fig:SNRROSATProfiles}[b]), the model underpredicts the observed intensity
by about a third ($\sim$30--$40 \times 10^{-6}~\rassrate$). For the middle data point, the
discrepancy is even larger, but this data point may be contaminated by two inaccurately removed
point sources (see \S\ref{subsec:SuzakuR45DiscrepancyDiscussion}, below). The disagreement between
the model and the observations shown in Figure~\ref{fig:SNRROSATProfiles}(b) is not due to a problem
with the SNR model itself, but instead is indicative of the general discrepancy between \suzaku\ and
the \rosat\ R45 band already noted in \S\ref{subsubsec:AdditionalDetails}.


\section{DISCUSSION}
\label{sec:Discussion}

The main goal of this project is to test the hypothesis that the bright arc in the 1/4~\kev\ SXRB
at $l \approx 247\degr$, $b \approx -64\degr$ is the edge of a bubble blown by an extraplanar
SN. In \S\ref{subsec:IsTheArcAnSNR} below, we will discuss
this SNR scenario. However, first we discuss the discrepancy between the \suzaku\ and \rosat\ R45
intensities, noted in \S\S\ref{subsubsec:AdditionalDetails} and \ref{subsec:SNRProfiles}.

\subsection{The Discrepancy between \suzaku\ and the \rosat\ R45 Band}
\label{subsec:SuzakuR45DiscrepancyDiscussion}

As was noted in \S\ref{subsubsec:AdditionalDetails}, during the course of our analysis we
discovered a discrepancy between our \suzaku\ spectra and the \rosat\ R45 (3/4~\kev)
count-rates. Table~\ref{tab:R45Countrates} compares the R45 count-rates predicted by our \suzaku\
fit results with the observed count-rates, averaged over circles of radius 0.5\degr\ and 1\degr. For
the R45 count-rates extracted from the 1\degr\ circles, the models underpredict the rates by
$\approx$$40 \times 10^{-6}~\rassrate$ for the arc-interior and arc-background directions, and by
$53 \times 10^{-6}~\rassrate$ for the on-arc direction. The on-arc discrepancy increases to
$99\times 10^{-6}~\rassrate$ when we use a 0.5\degr\ circle. The SNR model also underpredicts
the R45 count-rate (\S\ref{subsec:SNRProfiles}). The fact that the CIE and SNR models both
exhibit this discrepancy implies that the discrepancy is not due to a problem with the SNR model.

\begin{deluxetable}{lccc}
\tablewidth{0pt}
\tablecaption{Comparing Model and Observed R45 Count-rates\label{tab:R45Countrates}}
\tablehead{
		&			& \multicolumn{2}{c}{Observed}			\\
\cline{3-4}
Direction	& \colhead{Model}	& \colhead{0.5\degr}	& \colhead{1\degr}	\\
}
\startdata
Arc interior	& 75			& $109 \pm 9$		& $119 \pm 5$	\\
On arc		& 69			& $168 \pm 12$		& $122 \pm 5$	\\
Arc background	& 55			& $100 \pm 9$		& $95 \pm 5$	\\
\enddata
\tablecomments{All values are in $10^{-6}$~R45 \rassrate. The observed count-rates were averaged over circles of radius 0.5\degr\ and 1\degr.}
\end{deluxetable}

\citet{henley08a} also noticed a discrepancy between their \suzaku\ spectra and the corresponding \rosat\ count-rates.
They partially overcame this discrepancy by adding a \texttt{vphabs} absorption component to their
model. This extra component modeled contamination on the XIS1 optical blocking filter over and above
the contamination already included in the CALDB. \citet{henley08a} found that they needed an extra
carbon column density of $0.28 \times 10^{18}$~\pcmsq, in addition to the CALDB value of $3.1 \times
10^{18}$~\pcmsq\ at the center of the XIS1 chip. As the systematic uncertainty on the contamination
thickness is $\sim$$0.5 \times 10^{18}$~\pcmsq,\footnote{http://heasarc.gsfc.nasa.gov/docs/suzaku/analysis/xis0.html} this
correction is not unreasonable. However, we were unable to obtain a good fit simultaneously to the
\suzaku, R12, and R45 data for the on-arc direction, even with an extra \texttt{vphabs} absorption
component in our model.

\begin{figure*}
\plottwo{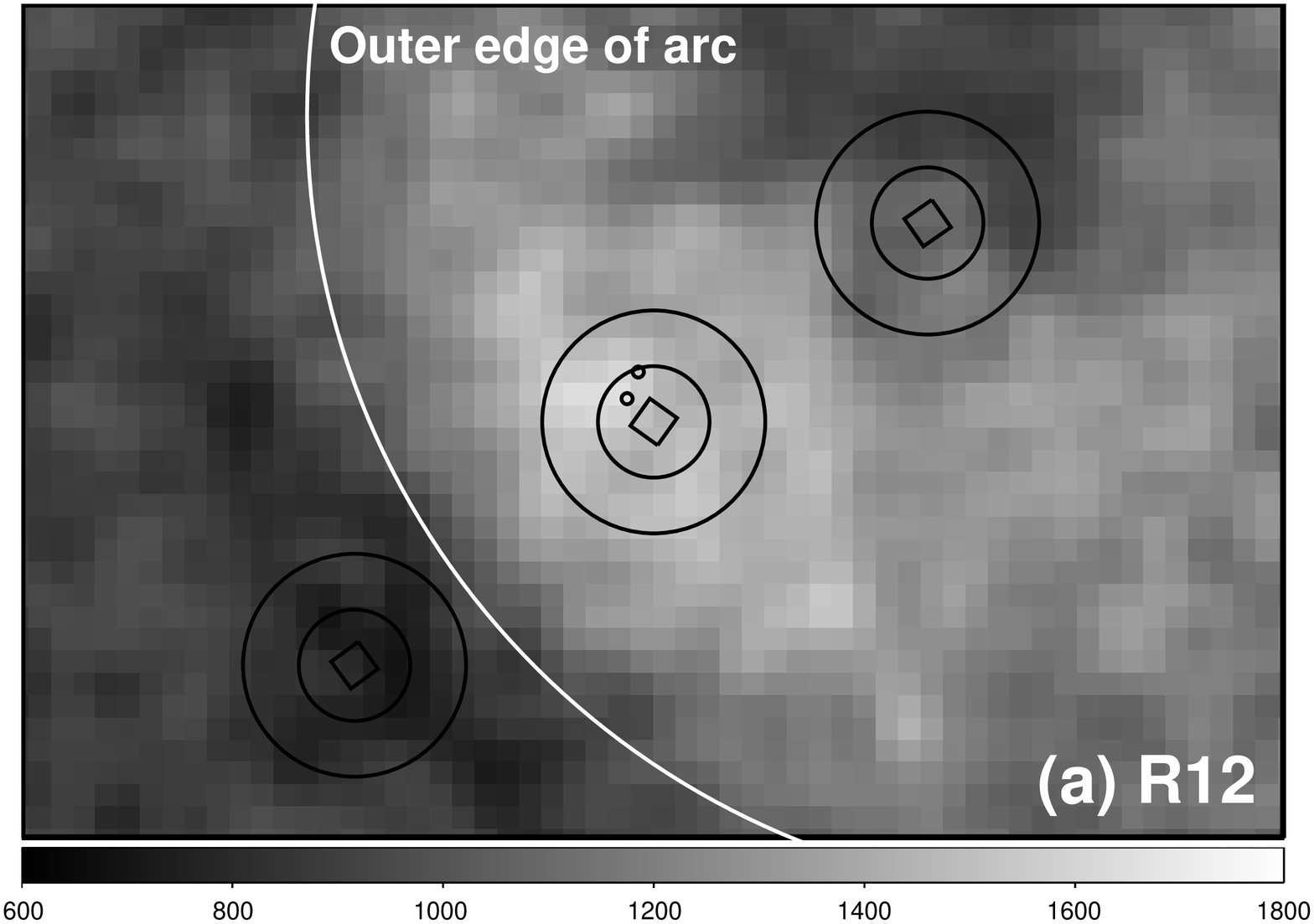}{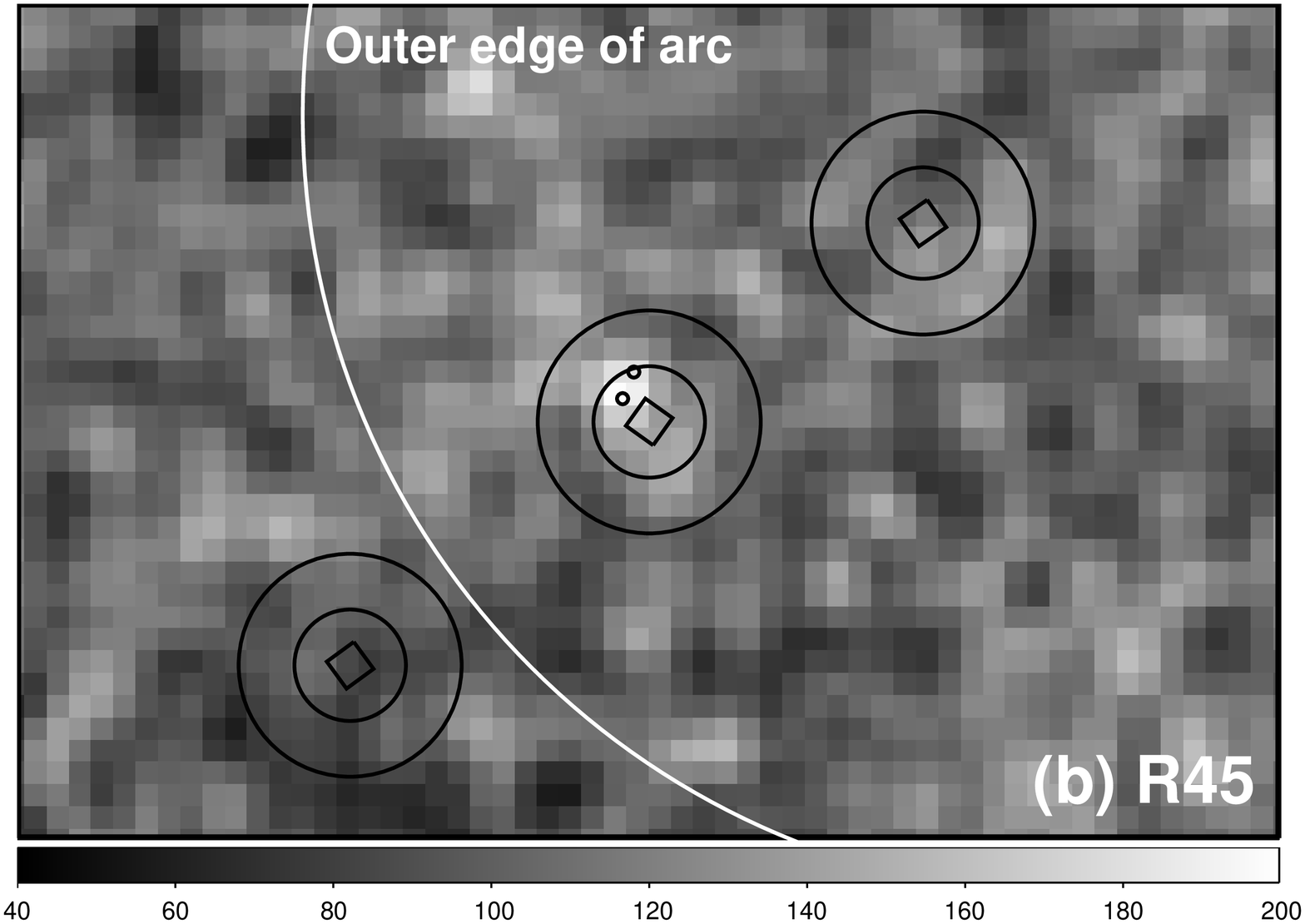}
\caption{Maps of the RASS (a) R12 and (b) R45 count-rate in the vicinity of the arc.
The data have been smoothed with a Gaussian whose standard deviation is 2 times the pixel size.
The units on the grayscale bar are $10^{-6}~\rassrate$.
The squares shows the approximate XIS fields of view for our \suzaku\ pointings
(top to bottom: arc interior, on arc, arc background).
The surrounding circles are of radius 0.5\degr\ and 1\degr.
The two smallest circles near the on-arc pointing show the sources 1RXS J023800.5$-$390505 and 1RXS J023734.5$-$391925, mentioned in the text.
The white line outlines the outer edge of the arc.
\label{fig:R45ArcImage}}
\end{figure*}

We think that the discrepancy may be partly due to the fact that the R45 emission is mottled on
scales of $\sim$1\degr\ (see Figure~\ref{fig:R45ArcImage}[b]). The observed R45 count-rates are
averaged over bright and faint mottled regions. For the on-arc direction, this includes a
particularly bright region, which may be partly due to inaccurate removal of the point sources 1RXS
J023800.5$-$390505 and 1RXS J023734.5$-$391925 (whose positions are shown in
Figure~\ref{fig:R45ArcImage}).  Because of the smoothing in Figure~\ref{fig:R45ArcImage}, and because
the size of the XIS field of view is similar to the RASS pixel size ($17.8\arcmin \times 17.8\arcmin$
versus $12\arcmin \times 12\arcmin$), one cannot accurately determine the R45 count-rate in the area
exactly corresponding to the XIS field of view. If our \suzaku\ pointings happen to be toward
fainter parts of the mottling, while the \rosat\ count-rates include both bright and faint parts of
the mottling, then the \rosat\ fluxes will be systematically brighter than the \suzaku\ fluxes.
In contrast, although the R12 data is much more variable over the whole sky than the R45 data,
Figure~\ref{fig:R45ArcImage}(a) shows that the R12 emission does not seem to be as mottled on
small angular scales.

The discrepancy may also be partially due to quiescent SWCX emission that is at a higher level in
the \rosat\ data than in the \suzaku\ data. Variations in the SXRB count-rate on a timescale of a
few days, referred to as ``long-term enhancements'' (LTEs), were removed from the RASS data
\citep{snowden95}. These LTEs are now thought to be due to variations in the heliospheric and/or
geocoronal SWCX emission \citep{cravens01}.  However, even after the LTEs have been removed, a
quiescent level of SWCX emission may remain in the data. In the R45 band, this SWCX emission would
be dominated by \OVII\ and \OVIII\ emission.  At high ecliptic latitudes, such as that of the arc
($\beta \sim -60\degr$), the heliospheric SWCX emission from these lines is expected to be brighter
at solar maximum (such as 1990/1991, when the RASS was carried out) than at solar minimum (when our
\suzaku\ observations were carried out) \citep{koutroumpa06}. \citet{koutroumpa07} give \OVII\ and
\OVIII\ intensities for various \chandra, \xmm, and \suzaku\ observations of the SXRB carried out at
solar maximum and solar minimum. Using the intensities predicted by their ``ground level'' model
(that is, excluding short-term solar wind enhancements) for the so-called ``southern Galactic filament''
(SGF), we estimate that the R45 count-rate at high ecliptic latitudes due to quiescent heliospheric
SWCX is $\sim$$10 \times 10^{-6}~\rassrate$ higher at solar maximum than at solar minimum. This
could explain $\sim$1/3 of the discrepancy between \suzaku\ and R45.

\citet{yoshino09} compared observed \rosat\ R45 count-rates with those predicted by \suzaku\
spectra from 14 different directions (1 direction, the North Ecliptic Pole, was observed twice).
They found that the \rosat\ intensities were systematically brighter than the \suzaku\ intensities.
For 5 of these directions, the \rosat\ count-rates may be significantly contaminated by LTEs. After
removing these 5 directions, \citet{yoshino09} found that the \rosat\ rates are an average of $17 \times
10^{-6}~\rassrate$ higher than the corresponding \suzaku\ rates. They found that much of this offset
could be due to a difference in the point source sensitivity between the two datasets. They also
considered variations in the heliospheric SWCX emission between the two datasets.

\begin{figure}
\plotone{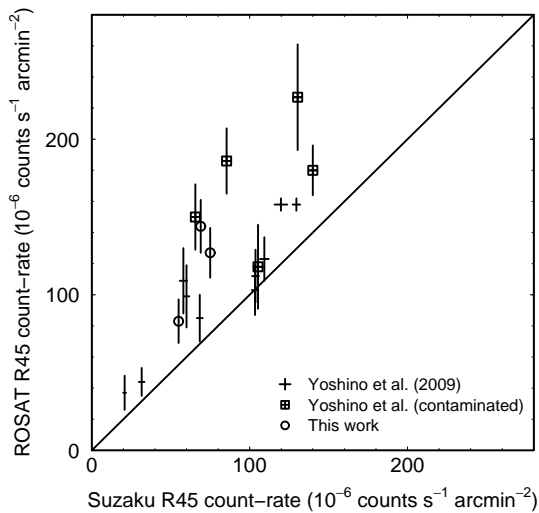}
\caption{Comparison of the R45 count-rates predicted by various \suzaku\ observations with the
observed \rosat\ R45 count-rates \citep[after][]{yoshino09}. The crosses show data from
\citeauthor{yoshino09}. The datapoints surrounded by squares are identified by
\citeauthor{yoshino09} as being from possibly contaminated regions of the RASS. Our arc data are
shown by open circles.  The diagonal solid line shows equality.\label{fig:R45vsSuzaku}}
\end{figure}

Figure~\ref{fig:R45vsSuzaku} shows a comparison of the observed \rosat\ R45 count-rates and
the rates predicted by \suzaku; the plot shows the results from \citet[see their Figure~6]{yoshino09}
and our arc results. For the purposes of this plot, we re-extracted the observed R45 count-rates
using a circle of radius 0.3\degr, to match \citet{yoshino09}. As can be seen, the discrepancy that
we see for our three spectra is not unusually large when compared with \citeauthor{yoshino09}'s
results.

Our foreground LB/SWCX model is derived from RASS R12 data \citep{snowden00}, and so our foreground
model should already include a reasonable estimate of the contamination of the R12 emission by solar
maximum heliospheric SWCX emission and unresolved point sources. This model does not make a major
contribution in the \suzaku\ band (see Figs.~\ref{fig:Spectra+Model} and
\ref{fig:Spectra+SNRModel}). In addition, as noted above, the R12 data do not seem to be mottled
on small angular scales as the R45 data (see Figure~\ref{fig:R45ArcImage}). We therefore think that
our combining the \suzaku\ and R12 data should yield reasonably accurate results.

\subsection{Is the Arc the Edge of a Supernova Remnant?}
\label{subsec:IsTheArcAnSNR}

In \S\ref{sec:SNRModels} we used a multicomponent model of the SXRB, which include emission from a
SNR, to analyze our three \suzaku\ + R12 spectra. In \S\ref{subsec:SNRResults} we showed that our
spectra and the observed size of the arc are reasonably well explained by a model in which the arc
is the bright edge of a $\sim$100,000-yr old SNR, blown by a SN explosion with
$E_0 = 0.5 \times 10^{51}~\erg$ at a height of $\sim$1--2~\kpc. (Despite its age, we consider such a
SNR to be young, as it is still in the adiabatic stage of its evolution, before the formation of a
cool shell.)

It is important to note that the SNR model described in \S\ref{sec:SNRModels} does not fit the
spectra better than the CIE models described in \S\ref{sec:CIEModels}. As the three fits in
Table~\ref{tab:FitResults} are completely independent, we can add the values of \chisq\ and
the degrees of freedom to give $\chisq = 1167$ for 1007 degrees of freedom for the CIE model.
In constrast, our preferred SNR models give $\chisq \approx 1220$--1250 for 1009 degrees of
freedom. (Note that our two models are not ``nested'', so we cannot use the $F$-test to measure
the significance of this difference in \chisq.) The SNR models appear to do slightly worse in
terms of \chisq, but we point out that the SNR scenario provides a good explanation
for the arc's general morphology.

It should also be noted that there are some discrepancies between the SNR models and the
observations: the SNR models are generally either a factor of $\sim$3--4 too bright or a factor of
$\sim$2 too small. In addition, the models do not match the observed width of the arc
(\S\ref{subsec:SNRProfiles}).  We will first discuss some possible explanations for these
discrepancies (\S\S\ref{subsubsec:BrightessAndSizeOfArc} and \ref{subsubsec:ArcMorphology}),
before discussing other possible observations of the arc (\S\ref{subsubsec:OtherObservations}).

\subsubsection{The Brightness and Size of the Arc}
\label{subsubsec:BrightessAndSizeOfArc}

At early epochs, when the model SNRs are still edge-brightened, their X-ray surface brightnesses and physical sizes
are most strongly affected by the ambient density, $n_0$. As the X-ray emission comes from shock-heated ambient medium,
a denser ambient medium gives brighter remnants, as well as hastening the formation of the dense shell.
A denser ambient medium also leads to smaller remnants, as in the Sedov phase the radius at a given age is proportional
to $n_0^{-1/5}$ \citep[e.g.][\S12.2b]{spitzer78}. The apparent size of an object also depends on its distance. In the
case of our SNR models, the distance is derived from the height, $z$, corresponding to the ambient density \citep{shelton06},
using a Galactic latitude of $-60\degr$.

Model 1800C ($E_0 = 0.5 \times 10^{51}~\erg$, $\Beff = 5.0~\microgauss$, $n_0 = 0.005~\pcc$) gives the
best result in terms of the SNR surface brightness -- of all our SNR models, the best-fitting normalization for 1800C
is the closest to 1 ($0.78^{+0.07}_{-0.09}$). However, this model gives a SNR radius that is a factor
of $\sim$2 too small. A simple way of resolving this discrepancy is to assume that the SNR is a
factor of $\sim$2 closer than expected, without changing the ambient density. The surface brightness
and physical size of the SNR would be unaffected, while the apparent size of the SNR would be
$\sim$2 times greater, and in better agreement with the observed size of the arc. This suggested
resolution requires the height at which $n_0 = 0.005~\pcc$ to be $\approx$1~\kpc, as opposed to
1.8~\kpc. Is a density of 0.005~\pcc\ at $z = 1~\kpc$ plausible?

The density model $n_0(z)$ used by \citet{shelton06} comes from \citet{ferriere98a}. At $z = 1~\kpc$, this model
gives $n_0(\mathrm{H} + \mathrm{He}) = 0.016~\pcc$, assuming 10\%\ He by number. This density is dominated by the
neutral medium (37\%) and the warm ionized medium (61\%), the remaining 2\%\ being due to the hot ionized medium.
\citeauthor{ferriere98a}'s model for the warm ionized medium comes from the distribution of free electrons derived
by \citet{cordes91}, while her model for the neutral medium comes from \citet{dickey90}.

We used a Monte Carlo method to investigate whether or not a density of 0.005~\pcc\ at $z = 1~\kpc$
was plausible.  For each trial, we randomly varied each parameter of the warm-ionized and neutral
models according to its error bar, and calculated $n_0(z = 1~\kpc)$ from the resulting density
model. \citet{dickey90} do not quote errors for the neutral model's parameters, so we assumed 50\%\
errors. We included the hot ionized medium in $n_0(z)$, but did not vary its model parameters. We
carried out 100,000 trials. Approximately 24\%\ of the trials were discarded because the random
number generator gave one or more negative (and hence unphysical) model parameters for those
trials. Of the remaining trials, 7\%\ gave a density $n_0(z=1~\kpc) \le 0.005~\pcc$. We therefore
cannot rule out the possibility that the density at $z = 1~\kpc$ is as low as 0.005~\pcc,
corresponding to model 1800C. As a result, we cannot rule out the possibility that the arc is
well-described by model 1800C at $z \sim 1~\kpc$ instead of $z = 1.8~\kpc$, although this model is
a less favored option.

The models with $E_0 = 0.5 \times 10^{51}~\erg$ for $z = 850~\pc$ (850A, 850B, 850C) are in better
agreement with the observed radius of the arc, but they are a factor of $\sim$3 too bright. As
stated above, the surface brightness of a model SNR depends on the density of the ambient
medium. However, as the emission is dominated by line emission from ionized metals, the SNR surface
brightness also depends upon the metallicity of that medium. If the metallicity of the interstellar
medium above the disk were a factor of $\sim$3 lower than our assumed value \citep{anders89}, the
model SNR surface brightness would be in better agreement with that of the arc.
We also note that uncertainties in the model line emissivities may be important, although
this is more difficult to quantify.

The halo gas-phase abundances of Si, Mg, and Fe, which are all important line emitters in the
1/4~\kev\ band, are $\sim$0.6, $\sim$0.3, and $\sim$0.2 solar \citep[using \citealt{anders89} as a
reference]{savage96}. However, these values do not imply a subsolar metallicity for the halo;
instead, these elements are assumed to be depleted on to dust. In contrast, S has a solar gas-phase
abundance in the halo \citep{savage96}, implying it is not depleted on to dust.  Also, the
abundance of Ne (which, being a noble gas, is not expected to be depleted) toward an X-ray binary
(4U~1820$-$303) at $z \approx 1~\kpc$ is $1.2 \pm 0.2$ solar \citep{yao06}. (However, this source
is at low Galactic latitude [$b = -7.91\degr$], so the sightline samples the disk as well as the halo.)

These measurements suggest that the halo has a solar metallicity, at least for $z \la 2~\kpc$. However,
the X-ray emission from a halo SNR depends upon the gas-phase metallicity in the shock-heated gas.
Whether or not this metallicity is subsolar depends on how quickly the dust is destroyed behind the
shock. In gas of temperature $T = 1 \times 10^6~\K$ and density $n$, the rate of change of a dust grain's
radius $a$ due to thermal sputtering is \citep{seab87}
\begin{equation}
	\frac{da}{dt} \sim 10^{-3} \left( \frac{n}{\pcc} \right)~\angstrom~\yr^{-1},
\end{equation}
and so the lifetime $\tau$ of a dust grain in the gas is
\begin{eqnarray}
	\tau	&\equiv&	\frac{a}{da/dt} 	\nonumber \\
		&\sim&		10^6 \left( \frac{a}{100~\angstrom} \right) \left( \frac{0.1~\pcc}{n} \right)~\yr.
\end{eqnarray}
The density $n = 0.1~\pcc$ used in the above expression is the approximate density in the immediate
post-shock region of the SNR models at $z = 850~\pc$; this region is where most of the X-ray
emission originates. The lifetime of a 100-\angstrom\ dust grain in such a SNR is an order of magnitude greater
than the SNR age given by our spectral analysis ($\sim$$10^5~\yr$), and dust grains larger than
100~\angstrom\ would survive even longer. We would therefore expect that the X-ray emission from a
young halo SNR would reflect the depleted gas-phase abundances in \citet{savage96}, not the total
halo abundances.

The effect that these depleted abundances would have on our model SNR spectra depends on the details of
which elements are depleted and by how much. However, as the gas-phase halo abundances of several
elements are $\sim$$1/3$ solar \citep{savage96}, it seems reasonable to suggest that our model SNR
spectra, calculated using \citet{anders89} abundances, may be a factor of $\sim$3 too bright. Using
a metallicity of $\sim$$1/3$ solar for a SNR with $E_0 = 0.5 \times 10^{51}~\erg$ at $z = 850~\pc$
($n_0 = 0.02~\pcc$) would bring the model SNR surface brightness into better agreement with that of
the arc. If the arc is a young extraplanar SNR whose X-ray--emitting gas has a metallicity of
$\sim$$1/3$ solar, our analysis would then favor a height $z \sim 1~\kpc$ for the arc, as a SNR at
greater height (lower ambient density) would have too low a surface brightness, as well as being
too small.

\subsubsection{The Morphology of the Arc}
\label{subsubsec:ArcMorphology}

As noted in \S\ref{subsec:SNRProfiles}, the observed radial width of the arc is larger than that
predicted by our SNR model. We investigated this discrepancy by looking for known Galactic
SNRs at an analogous stage in their evolution (i.e. roughly half way to the formation of the cool
shell), in order to see how their observed X-ray morphologies compared with our model.

We carried out a SNR simulation with $E_0 = 0.5 \times 10^{51}~\erg$, $\Beff = 0$
(model type A), with an ambient density ($n_0 = 2~\pcc$) representative of the disk, rather than the
halo. From this simulation, we estimated that a $\sim$15,000-yr old SNR in the disk would be at a
similar stage in its evolution as our best-fitting halo SNR model for the arc. We found that the
Vela SNR (G263.9$-$3.3) provides a good analog to the arc: based on the apparent origin of
X-ray--emitting explosion fragments beyond the blastwave and the proper motion of the Vela pulsar
(PSR~B0833$-$45), \citet{aschenbach95} estimated the age of the remnant to be $18,000 \pm 9,000$~yr,
consistent with the pulsar's spin-down age of $11,000$~yr \citep{taylor93}.

As with the analogous models of halo SNRs, the new simulation with $n_0 = 2~\pcc$ predicts that
a SNR of the age of Vela would have an X-ray--bright rim whose width is $\sim$1/10 of
the radius of the SNR (cf. Figs.~\ref{fig:SNRModelProfiles} and \ref{fig:SNRROSATProfiles}).
In contrast, Vela exhibits an asymmetrical bright rim to the north and east whose width is
$\sim$half the remnant radius \citep{aschenbach95}.

Our SNR models underpredict the widths of the arc and of the Vela SNR's X-ray--bright rim.  These
discrepancies may be due to the limitations of our SNR simulations, which are one-dimensional, and
assume a uniform ambient density and a uniform non-thermal pressure in the radial direction.
Three-dimensional simulations could include non-uniform ambient densities and different
magnetic field geometries, which could affect the predicted X-ray morphology. Such simulations could
also simulate hydrodynamical instabilities, which would broaden the apparent width of the
arc. E.~Raley (2009, private communication) has provided us with the results of a 3-D simulation of
a SNR with $E_0 = 0.5 \times 10^{51}~\erg$ evolving in zero magnetic field at $z = 400~\pc$ (see
also \citealt{raley07}). Using these results, K.~Kwak (2009, private communication) has provided us
with spectra calculated for various sightlines across the remnant, assuming CIE. For a young
remnant, the 3-D model does indeed predict a broader X-ray--bright rim than the 1-D model. However,
the 1-D SNR models that we used include self-consistent modeling of the ionization evolution, which
we used to calculate non-equilibrium X-ray spectra. Developing a 3-D SNR model that includes
ionization evolution is beyond the scope of this paper.

We conclude this discussion of the arc's morphology by noting that the arc does not trace a full
circle. Such asymmetries are not uncommon in SNRs (see, e.g., the collection of SNR images at the \rosat\
Guest Observer Facility\footnote{http://heasarc.nasa.gov/docs/rosat/gallery/snrs.html} and the
\chandra\ Supernova Remnant Catalog\footnote{http://hea-www.cfa.harvard.edu/ChandraSNR/}). The arc's
asymmetry may be due to a non-uniform ambient medium and/or a complicated magnetic field geometry,
both of which are beyond the reach of our 1-D simulations.

\subsubsection{Non-X-ray Observations, and Future X-ray Observations}
\label{subsubsec:OtherObservations}

We have considered further ways in which we could test the hypothesis that the arc is the edge of an extraplanar
SNR. Our X-ray spectral analysis implies that, if the arc is a SNR, then it is young, in the sense that it has
not yet formed a dense cool shell. This is unfortunate, as it means we do not expect \Halpha\ emission due
to recombination in the cooling shell, nor \HI\ emission from the gas that has already cooled. The
Southern \Halpha\ Sky Survey Atlas (SHASSA; \citealp{gaustad01,finkbeiner03}) shows no sign of an \Halpha-emitting
shell around the arc.

The hydrodynamical SNR models that we have used in our spectral analysis also make predictions for
the column densities of various highly ionized metals (e.g., \NV, \OVI, \OVII, \OVIII). In
principle, therefore, far-ultraviolet and X-ray absorption line spectroscopy could be used to test
the results of our spectral analysis. Unfortunately, in practice it would be extremely difficult to
detect unambiguously an enhanced ion column density due to an extraplanar SNR, given the sparsity
of stars in this region.

Table~\ref{tab:ColumnDensities} shows various ion column densities predicted for a 100,000-yr old
SNR at $z \ge 850~\pc$ with $E_0 = 0.5 \times 10^{51}~\erg$.  The predicted column density is
dependent upon $z$ and \Beff: models at lower $z$ (and hence higher ambient density) or with larger
\Beff\ give larger column densities. Table~\ref{tab:ColumnDensities} shows the range of values
predicted by the various models. The \NV\ and \OVI\ columns (and, to a lesser extent, the \OVII\
column) are peaked at the edge of the remnant (see Figs. 8 and 9 in \citealt{shelton98}; that model
corresponds to our model 1300B). For these ions, Table~\ref{tab:ColumnDensities} shows both the peak
column, and the column in the SNR interior. The \OVIII\ column does not peak at the edge of the SNR.

\begin{deluxetable*}{lcccccc}
\tablewidth{0pt}
\tablecaption{Predicted SNR Ion Column Densities and Observed Ion Column Densities\label{tab:ColumnDensities}}
\tablehead{
		& \multicolumn{2}{c}{Predicted SNR Column Density}		&& \multicolumn{3}{c}{Observations}						\\
\cline{2-3} \cline{5-7}
		& \colhead{Peak}		& \colhead{Interior}		&& \colhead{Column Density}	& \colhead{Number of}	&			\\
\colhead{Ion}	& \colhead{($10^{14}~\pcmsq$)}	& \colhead{($10^{14}~\pcmsq$)}	&& \colhead{($10^{14}~\pcmsq$)}	& \colhead{Sightlines}	& \colhead{Ref.}	\\
}
\startdata
\NV		& 0.2--1.6			& 0.03--0.19			&& $0.57 \pm 0.36$\tablenotemark{a}& 9			& 1			\\
\OVI		& 3.9--14.2			& 0.6--2.1			&& $2.5 \pm 1.0$\tablenotemark{a}& 91			& 2			\\
\OVII		& 13--43			& 10--23			&& $\sim$100--200		& 3			& 3,4,5			\\
\OVIII		& \nodata			& 1.3--8.5			&& $\sim$10--100		& 2			& 3,4			\\
\enddata		
\tablerefs{(1) \citealt{sembach92}; (2) \citealt{savage03}; (3) \citealt{yao07a}; (4) \citealt{yao07b}; (5) \citealt{yao08}.}
\tablecomments{The predicted column densities are from 100,000-yr old SNRs at $z \ge 850~\pc$ with $E_0 = 0.5 \times 10^{51}~\erg$.
The observed values indicate the variation in the column densities detected from multiple sightlines over the sky.}
\tablenotetext{a}{Mean $\pm$ standard deviation}
\end{deluxetable*}

Table~\ref{tab:ColumnDensities} also shows the range of column densities observed in multiple
directions across the sky. In general, the column densities predicted by the SNR are smaller than
the spread in the observed column densities, and so an increased column density in the vicinity of
the arc could not be unambiguously attributed to an extraplanar SNR. Furthermore, the predicted
\OVII\ columns are of the same order of magnitude as the errors on the \OVII\ measurements.
Some of the models (with $z = 850~\pc$) predict peak \NV\ and \OVI\ column densities near the
remnant edge that are significantly larger than the variation across the sky. However, observing
these large column densities would require a fortuitously positioned background source. In general,
using absorption line spectroscopy to test whether or not the arc is the edge of an extraplanar SNR
would require column density measurements from several different sightlines on and around the arc.

Perhaps the best way of further testing the hypothesis that the arc is the edge of an extraplanar
SNR will be with future, higher-resolution X-ray spectrometers. Although our \suzaku\ spectra are
consistent with CIE models, it is possible that higher resolution spectra could reveal unambiguous
signs of non-equilibrium ionization that we would expect behind the blastwave of a young extraplanar
SNR.  Sensitivity in the 1/4 \kev\ band, where the arc is bright, but where we currently only have
the low-resolution \rosat\ R12 data, would be particularly useful.

\subsection{Other extraplanar SNRs?}

\citet{shelton07} and \citet{lei09} have analyzed the halo \OVI\ and X-ray emission in the direction
$l \approx 279\degr$, $b \approx -47\degr$, which samples an X-ray--bright region visible
in the 1/4-\kev\ RASS maps. They used a nearby ($d = 230~\pc$) shadowing
filament to separate the foreground (LB/SWCX) and background (halo) emission. This filament is
visible toward the upper-right corner of Figure~\ref{fig:ArcImage}.

\citet{shelton07} showed that the \OVI-to-1/4 \kev\ emission ratio was a good age diagnostic for an
extraplanar SNR. For the halo beyond the shadowing filament, they found that this ratio
was consistent with that from a $\sim$40,000--70,000-yr old SNR with $n_0 = 0.01~\pcc$,
$E_0 = 0.5 \times 10^{51}~\erg$, $\Beff = 2.5~\microgauss$ (model 1300B in
Table~\ref{tab:ModelParameters}). More generally, they found that if the hot gas that they observed
can be compared to that in an undisturbed extraplanar SNR, the time since heating is
$\sim$$10^4$--$10^5~\yr$.

Using a combination of \suzaku, \rosat\ R12, and \fuse\ \OVI\ data, \citet{lei09} constructed a
differential emission measure (DEM) model for the halo beyond the shadowing filament.  Their DEM is
a broken power-law between $T \sim 10^5~\K$ and $T \sim 10^7~\K$, with a break at $T_\mathrm{break}
\sim 10^6~\K$. They suggest that the lower-temperature part of the DEM ($T < T_\mathrm{break}$) is
due to an extraplanar SNR. The \OVI\ and soft X-ray intensities of this lower-temperature component
are best matched by a $\sim$180,000-yr old SNR with $n_0 = 0.02~\pcc$ ($z = 850~\pc$). Given the
uncertainties in the modeling, \citet{shelton07} and \citet{lei09} obtained consistent results:
both studies found that the \OVI\ and 1/4 \kev\ X-ray emission from the halo beyond the shadowing filament
are consistent with the emission from a pre--shell-formation SNR at $z \sim 1~\kpc$.

\citet{shelton06} predicted that there would be approximately one bright, pre--shell-formation SNR
per hemisphere at high-latitudes. Our arc analysis and the analysis of the filament region
\citep{shelton07,lei09} suggest the presence of two halo SNRs at similar heights and of similar ages
in the southern Galactic hemisphere. Given the uncertainties in the computer simulations and
SN rate, this result should not be considered a refutation of \citepossessive{shelton06}
prediction. However, given this result, we expect that most other directions in the southern
Galactic hemisphere would not exhibit the properties of young, extraplanar SNRs. This could be
tested by measuring the \OVI-to-1/4 \kev\ ratio for a large number of high-latitude directions.


\section{SUMMARY}
\label{sec:Summary}

We have analyzed a set of three \suzaku\ spectra obtained from pointings on and around a bright arc
that appears in the 1/4~\kev\ soft X-ray background at $l \approx 247\degr$, $b \approx -64\degr$,
and have tested the hypothesis that the arc is the edge of an extraplanar SNR. For this purpose we
have used spectral models generated from 1-D hydrodynamical simulations of SNRs at a variety of
heights above the disk. We also supplemented our \suzaku\ spectra at lower energies with \rosat\
R12 (1/4 keV) data.

The three \suzaku\ + R12 spectra and the observed size of the arc are reasonably well explained by a
model in which the arc is the bright edge of a $\sim$100,000-yr old SNR at a height
$z \sim 1$--2~\kpc\ (ambient density $n_0 = 0.02$--0.005~\pcc), blown by a supernova with
$E_0 = 0.5 \times 10^{51}~\erg$. The remnant is still in the adiabatic stage of its evolution,
having not yet formed a cool shell.

The agreement between the model and the observations is not perfect: the models are generally either
a few times too bright, or a factor of $\sim$2 too small. The agreement between the model and the
observations can be improved if the metallicity of the X-ray--emitting gas is $\sim$1/3 solar.  Such
a subsolar metallicity is plausible, as metals depleted on to dust are unlikely to have been
returned to the gas phase within the $\sim$100,000-yr lifetime of the remnant. If the metallicity is
$\sim$1/3 solar, this scenario would favor $z \sim 1$~\kpc\ for the arc, as higher remnants would be
too faint and too small.

The radial width of the arc is underpredicted by our SNR model. The Vela SNR, a known remnant in the
Galactic disk, appears to be at a similar stage in its evolution as our arc SNR model. Like the arc,
its X-ray--bright rim is much wider than that predicted by our simulations. We suggest that this
discrepancy could be due to the 1-D nature of the simulations, as well as assumptions of uniform
ambient density and non-thermal pressure. Higher-dimensional hydrodynamical simulations may
better explain the morphology of the arc, but are beyond the scope of this study.

It should be noted that we have not tested other scenarios for the arc's formation, and also that
equilibrium models provide good fits to the \suzaku\ spectra. However, we conclude by echoing
\citet{shelton06}, and note that the extraplanar SNR scenario discussed here provides a good explanation
for an arc-shaped enhancement in the high-latitude soft X-ray background. If this scenario is
correct, it supports the idea that extraplanar supernovae contribute to the heating of the hot
halo gas.


\acknowledgements
We would like to thank Steve Snowden for helpful comments regarding HEASARC's X-ray Background Tool,
Liz Raley for helpful discussions on fitting the SNR spectral models to the data and for supplying
the 3-D SNR simulation mentioned in \S\ref{subsubsec:ArcMorphology}, and Kyujin Kwak for calculating
X-ray spectra from said simulation. We also thank the referee, whose comments have helped improve
this paper. This research has made use of data obtained from the \suzaku\
satellite, a collaborative mission between the space agencies of Japan (JAXA) and the USA
(NASA). The OMNI data were obtained from the GSFC/SPDF OMNIWeb interface at
http://omniweb.gsfc.nasa.gov. This research was funded by NASA grants NNH06ZDA001B-SUZ2 and
NNX08AZ83G, awarded through the \suzaku\ Guest Observer Program.

\bibliography{references}

\begin{thebibliography}{60}
\expandafter\ifx\csname natexlab\endcsname\relax\def\natexlab#1{#1}\fi

\bibitem[{Anders \& Grevesse(1989)}]{anders89}
Anders, E., \& Grevesse, N. 1989, Geochim. Cosmochim. Acta, 53, 197

\bibitem[{Arnaud(1996)}]{arnaud96}
Arnaud, K.~A. 1996, in ASP Conf. Ser. 101: Astronomical Data Analysis Software
  and Systems V, ed. G.~H. Jacoby \& J.~Barnes, 17

\bibitem[{Aschenbach {et~al.}(1995)Aschenbach, Egger, \&
  Tr{\"u}mper}]{aschenbach95}
Aschenbach, B., Egger, R., \& Tr{\"u}mper, J. 1995, Nature, 373, 587

\bibitem[{Ba{\l}uci{\'n}ska-Church \& McCammon(1992)}]{balucinska92}
Ba{\l}uci{\'n}ska-Church, M., \& McCammon, D. 1992, ApJ, 400, 699

\bibitem[{Breitschwerdt \& Schmutzler(1994)}]{breitschwerdt94}
Breitschwerdt, D., \& Schmutzler, T. 1994, Nature, 371, 774

\bibitem[{Burrows \& Mendenhall(1991)}]{burrows91}
Burrows, D.~N., \& Mendenhall, J.~A. 1991, Nature, 351, 629

\bibitem[{Chen {et~al.}(1997)Chen, Fabian, \& Gendreau}]{chen97}
Chen, L.-W., Fabian, A.~C., \& Gendreau, K.~C. 1997, MNRAS, 285, 449

\bibitem[{Cordes {et~al.}(1991)Cordes, Weisberg, Frail, Spangler, \&
  Ryan}]{cordes91}
Cordes, J.~M., Weisberg, J.~M., Frail, D.~A., Spangler, S.~A., \& Ryan, M.
  1991, Nature, 354, 121

\bibitem[{Cravens(2000)}]{cravens00}
Cravens, T.~E. 2000, ApJ, 532, L153

\bibitem[{Cravens {et~al.}(2001)Cravens, Robertson, \& Snowden}]{cravens01}
Cravens, T.~E., Robertson, I.~P., \& Snowden, S.~L. 2001, JGR, 106 (A11), 24883

\bibitem[{Dickey \& Lockman(1990)}]{dickey90}
Dickey, J.~M., \& Lockman, F.~J. 1990, ARA\&A, 28, 215

\bibitem[{Ferri{\`e}re(1998)}]{ferriere98a}
Ferri{\`e}re, K. 1998, ApJ, 497, 759

\bibitem[{Finkbeiner(2003)}]{finkbeiner03}
Finkbeiner, D.~P. 2003, ApJS, 146, 407

\bibitem[{Fujimoto {et~al.}(2007)Fujimoto, Mitsuda, McCammon, Takei, Bauer,
  Ishisaki, Porter, Yamaguchi, Hayashida, \& Yamasaki}]{fujimoto07}
Fujimoto,  R., et al. 2007, PASJ,   59, S133

\bibitem[{Galeazzi {et~al.}(2007)Galeazzi, Gupta, Covey, \&
  Ursino}]{galeazzi07}
Galeazzi, M., Gupta, A., Covey, K., \& Ursino, E. 2007, ApJ, 658, 1081

\bibitem[{Gaustad {et~al.}(2001)Gaustad, McCullough, Rosing, \&
  Van~Buren}]{gaustad01}
Gaustad, J.~E., McCullough, P.~R., Rosing, W., \& Van~Buren, D. 2001, PASP,
  113, 1326

\bibitem[{Henley \& Shelton(2008)}]{henley08a}
Henley, D.~B., \& Shelton, R.~L. 2008, ApJ, 676, 335

\bibitem[{Ishisaki {et~al.}(2007)Ishisaki, Maeda, Fujimoto, Ozaki, Ebisawa,
  Takahashi, Ueda, Ogasaka, Ptak, Mukai, Hamaguchi, Hirayama, Kotani, Kubo,
  Shibata, Ebara, Furuzawa, Iizuka, Inoue, Mori, Okada, Yokoyama, Matsumoto,
  Nakajima, Yamaguchi, Anabuki, Tawa, Nagai, Katsuda, Hayashida, Bamba, Miller,
  Sato, \& Yamasaki}]{ishisaki07}
Ishisaki,  Y., et al. 2007, PASJ, 59, S113

\bibitem[{Kalberla {et~al.}(2005)Kalberla, Burton, Hartmann, Arnal, Bajaja,
  Morras, \& P{\"o}ppel}]{kalberla05}
Kalberla, P.~M.~W., Burton, W.~B., Hartmann, D., Arnal, E.~M., Bajaja, E.,
  Morras, R., \& P{\"o}ppel, W.~G.~L. 2005, A\&A, 440, 775

\bibitem[{Koutroumpa {et~al.}(2007)Koutroumpa, Acero, Lallement, Ballet, \&
  Kharchenko}]{koutroumpa07}
Koutroumpa, D., Acero, F., Lallement, R., Ballet, J., \& Kharchenko, V. 2007,
  A\&A, 475, 901

\bibitem[{Koutroumpa {et~al.}(2006)Koutroumpa, Lallement, Kharchenko, Dalgarno,
  Pepino, Izmodenov, \& Qu{\'e}merais}]{koutroumpa06}
Koutroumpa, D., Lallement, R., Kharchenko, V., Dalgarno, A., Pepino, R.,
  Izmodenov, V., \& Qu{\'e}merais, E. 2006, A\&A, 460, 289

\bibitem[{Koyama {et~al.}(2007)Koyama, Tsunemi, Dotani, Bautz, Hayashida,
  Tsuru, Matsumoto, Ogawara, Ricker, Doty, Kissel, Foster, Nakajima, Yamaguchi,
  Mori, Sakano, Hamaguchi, Nishiuchi, Miyata, Torii, Namiki, Katsuda, Matsuura,
  Miyauchi, Anabuki, Tawa, Ozaki, Murakami, Maeda, Ichikawa, Prigozhin,
  Boughan, LaMarr, Miller, Burke, Gregory, Pillsbury, Bamba, Hiraga, Senda,
  Katayama, Kitamoto, Tsujimoto, Kohmura, Tsuboi, \& Awaki}]{koyama07}
Koyama,  K., et al. 2007, PASJ, 59, S23

\bibitem[{Kuntz \& Snowden(2000)}]{kuntz00}
Kuntz, K.~D., \& Snowden, S.~L. 2000, ApJ, 543, 195

\bibitem[{Lei {et~al.}(2009)Lei, Shelton, \& Henley}]{lei09}
Lei, S., Shelton, R.~L., \& Henley, D.~B. 2009, ApJ, in press
  (arXiv:0906.1532v1)

\bibitem[{Mitsuda {et~al.}(2007)Mitsuda, Bautz, Inoue, Kelley, Koyama, Kunieda,
  Makishima, Ogawara, Petre, Takahashi, Tsunemi, White, Anabuki, Angelini,
  Arnaud, Awaki, Bamba, Boyce, Brown, Chan, Cottam, Dotani, Doty, Ebisawa,
  Ezoe, Fabian, Figueroa, Fujimoto, Fukazawa, Furusho, Furuzawa, Gendreau,
  Griffiths, Haba, Hamaguchi, Harrus, Hasinger, Hatsukade, Hayashida, Henry,
  Hiraga, Holt, Hornschemeier, Hughes, Hwang, Ishida, Ishisaki, Isobe, Itoh,
  Iyomoto, Kahn, Kamae, Katagiri, Kataoka, Katayama, Kawai, Kilbourne,
  Kinugasa, Kissel, Kitamoto, Kohama, Kohmura, Kokubun, Kotani, Kotoku, Kubota,
  Madejski, Maeda, Makino, Markowitz, Matsumoto, Matsumoto, Matsuoka,
  Matsushita, McCammon, Mihara, Misaki, Miyata, Mizuno, Mori, Mori, Morii,
  Moseley, Mukai, Murakami, Murakami, Mushotzky, Nagase, Namiki, Negoro,
  Nakazawa, Nousek, Okajima, Ogasaka, Ohashi, Oshima, Ota, Ozaki, Ozawa,
  Parmar, Pence, Porter, Reeves, Ricker, Sakurai, Sanders, Senda, Serlemitsos,
  Shibata, Soong, Smith, Suzuki, Szymkowiak, Takahashi, Tamagawa, Tamura,
  Tamura, Tanaka, Tashiro, Tawara, Terada, Terashima, Tomida, Torii, Tsuboi,
  Tsujimoto, Tsuru, Turner, Ueda, Ueno, Ueno, Uno, Urata, Watanabe, Yamamoto,
  Yamaoka, Yamasaki, Yamashita, Yamauchi, Yamauchi, Yaqoob, Yonetoku, \&
  Yoshida}]{mitsuda07}
Mitsuda,  K., et al. 2007,   PASJ, 59, S1

\bibitem[{Norman \& Ikeuchi(1989)}]{norman89}
Norman, C.~A., \& Ikeuchi, S. 1989, ApJ, 345, 372

\bibitem[{Raley {et~al.}(2007)Raley, Shelton, \& Plewa}]{raley07}
Raley, E.~A., Shelton, R.~L., \& Plewa, T. 2007, ApJ, 661, 222

\bibitem[{Rasmussen {et~al.}(2009)Rasmussen, Sommer-Larsen, Pedersen, Toft,
  Benson, Bower, \& Grove}]{rasmussen09}
Rasmussen, J., Sommer-Larsen, J., Pedersen, K., Toft, S., Benson, A., Bower,
  R.~G., \& Grove, L.~F. 2009, ApJ, 697, 79

\bibitem[{Raymond \& Smith(1977)}]{raymond77}
Raymond, J.~C., \& Smith, B.~W. 1977, ApJS, 35, 419

\bibitem[{Robertson \& Cravens(2003{\natexlab{a}})}]{robertson03a}
Robertson, I.~P., \& Cravens, T.~E. 2003{\natexlab{a}}, JGR, 108 (A10), 8031

\bibitem[{Robertson \& Cravens(2003{\natexlab{b}})}]{robertson03b}
---. 2003{\natexlab{b}}, GeoRL, 30(8), 1439

\bibitem[{Savage \& Sembach(1996)}]{savage96}
Savage, B.~D., \& Sembach, K.~R. 1996, ARA\&A, 34, 279

\bibitem[{Savage {et~al.}(2003)Savage, Sembach, Wakker, Richter, Meade,
  Jenkins, Shull, Moos, \& Sonneborn}]{savage03}
Savage,  B.~D., et al. 2003, ApJS, 146, 125

\bibitem[{Schlegel {et~al.}(1998)Schlegel, Finkbeiner, \& Davis}]{schlegel98}
Schlegel, D.~J., Finkbeiner, D.~P., \& Davis, M. 1998, ApJ, 500, 525

\bibitem[{Seab(1987)}]{seab87}
Seab, C.~G. 1987, in Interstellar Processes, ed. D.~J. Hollenbach \& H.~A.
  Thronson, Jr. (Dordrecht: Reidel), 491

\bibitem[{Sembach \& Savage(1992)}]{sembach92}
Sembach, K.~R., \& Savage, B.~D. 1992, ApJS, 83, 147

\bibitem[{Shapiro \& Field(1976)}]{shapiro76}
Shapiro, P.~R., \& Field, G.~B. 1976, ApJ, 205, 762

\bibitem[{Shelton(1998)}]{shelton98}
Shelton, R.~L. 1998, ApJ, 504, 785

\bibitem[{Shelton(1999)}]{shelton99}
---. 1999, ApJ, 521, 217

\bibitem[{Shelton(2006)}]{shelton06}
---. 2006, ApJ, 638, 206

\bibitem[{Shelton {et~al.}(2007)Shelton, Sallmen, \& Jenkins}]{shelton07}
Shelton, R.~L., Sallmen, S.~M., \& Jenkins, E.~B. 2007, ApJ, 659, 365

\bibitem[{Smith {et~al.}(2007)Smith, Bautz, Edgar, Fujimoto, Hamaguchi, Hughes,
  Ishida, Kelley, Kilbourne, Kuntz, McCammon, Miller, Mitsuda, Mukai,
  Plucinsky, Porter, Snowden, Takei, Terada, Tsuboi, \& Yamasaki}]{smith07a}
Smith,  R.~K., et al. 2007, PASJ, 59,   S141

\bibitem[{Smith {et~al.}(2001)Smith, Brickhouse, Liedahl, \&
  Raymond}]{smith01a}
Smith, R.~K., Brickhouse, N.~S., Liedahl, D.~A., \& Raymond, J.~C. 2001, ApJ,
  556, L91

\bibitem[{Snowden {et~al.}(2004)Snowden, Collier, \& Kuntz}]{snowden04}
Snowden, S.~L., Collier, M.~R., \& Kuntz, K.~D. 2004, ApJ, 610, 1182

\bibitem[{Snowden {et~al.}(1998)Snowden, Egger, Finkbeiner, Freyberg, \&
  Plucinsky}]{snowden98}
Snowden, S.~L., Egger, R., Finkbeiner, D.~P., Freyberg, M.~J., \& Plucinsky,
  P.~P. 1998, ApJ, 493, 715

\bibitem[{Snowden {et~al.}(1997)Snowden, Egger, Freyberg, McCammon, Plucinsky,
  Sanders, Schmitt, Tr{\"u}mper, \& Voges}]{snowden97}
Snowden,  S.~L., et al. 1997,   ApJ, 485, 125

\bibitem[{Snowden {et~al.}(2000)Snowden, Freyberg, Kuntz, \&
  Sanders}]{snowden00}
Snowden, S.~L., Freyberg, M.~J., Kuntz, K.~D., \& Sanders, W.~T. 2000, ApJS,
  128, 171

\bibitem[{Snowden {et~al.}(1995)Snowden, Freyberg, Plucinsky, Schmitt,
  Tr{\"u}mper, Voges, Edgar, McCammon, \& Sanders}]{snowden95}
Snowden,  S.~L., et al. 1995, ApJ, 454, 643

\bibitem[{Snowden {et~al.}(1991)Snowden, Mebold, Hirth, Herbstmeier, \&
  Schmitt}]{snowden91}
Snowden, S.~L., Mebold, U., Hirth, W., Herbstmeier, U., \& Schmitt, J.~H.~M.~M.
  1991, Science, 252, 1529

\bibitem[{Spitzer(1978)}]{spitzer78}
Spitzer, L. 1978, Physical Process in the Interstellar Medium (New York: Wiley)

\bibitem[{Taylor {et~al.}(1993)Taylor, Manchester, \& Lyne}]{taylor93}
Taylor, J.~H., Manchester, R.~N., \& Lyne, A.~G. 1993, ApJS, 88, 529

\bibitem[{Toft {et~al.}(2002)Toft, Rasmussen, Sommer-Larsen, \&
  Pedersen}]{toft02}
Toft, S., Rasmussen, J., Sommer-Larsen, J., \& Pedersen, K. 2002, MNRAS, 335,
  799

\bibitem[{Winkler {et~al.}(2003)Winkler, Gupta, \& Long}]{winkler03}
Winkler, P.~F., Gupta, G., \& Long, K.~S. 2003, ApJ, 585, 324

\bibitem[{Yan {et~al.}(1998)Yan, Sadeghpour, \& Dalgarno}]{yan98}
Yan, M., Sadeghpour, H.~R., \& Dalgarno, A. 1998, ApJ, 496, 1044

\bibitem[{Yao {et~al.}(2008)Yao, Nowak, Wang, Schulz, \& Canizares}]{yao08}
Yao, Y., Nowak, M.~A., Wang, Q.~D., Schulz, N.~S., \& Canizares, C.~R. 2008,
  ApJ, 672, L21

\bibitem[{Yao \& Wang(2005)}]{yao05}
Yao, Y., \& Wang, Q.~D. 2005, ApJ, 624, 751

\bibitem[{Yao \& Wang(2006)}]{yao06}
---. 2006, ApJ, 641, 930

\bibitem[{Yao \& Wang(2007{\natexlab{a}})}]{yao07a}
---. 2007{\natexlab{a}}, ApJ, 658, 1088

\bibitem[{Yao \& Wang(2007{\natexlab{b}})}]{yao07b}
---. 2007{\natexlab{b}}, ApJ, 666, 242

\bibitem[{Yoshino {et~al.}(2009)Yoshino, Mitsuda, Yamasaki, Takei, Hagihara,
  Masui, Bauer, McCammon, Fujimoto, Wang, \& Yao}]{yoshino09}
Yoshino,  T., et al. 2009, PASJ,   submitted (arXiv:0903.2981v1)

\end{thebibliography}


\end{document}